\documentclass[a4paper,12pt,oneside]{article}
\usepackage[latin1]{inputenc}
\usepackage{amsmath}
\usepackage{amsfonts}
\usepackage{amssymb}
\usepackage{graphicx}
\usepackage{footnote}
\usepackage{float}
\usepackage{subfigure}
\DeclareGraphicsExtensions{.bmp,.png,.pdf,.jpg,.eps}

\usepackage[colorlinks]{hyperref}
\hypersetup{
    colorlinks=true,
    linkcolor=blue,
    filecolor=magenta,  
    urlcolor=[rgb]{0.00,0.00,0.50},    
    citecolor=red,
    }

\usepackage{url}

\advance \textheight by \topskip
\usepackage{amsmath}
\usepackage[english]{babel}
\makeatletter
 \@addtoreset{equation}{section}
\makeatother
\usepackage[latin1]{inputenc}
\usepackage{amsmath}
\numberwithin{equation}{section}

\advance \textheight by \topskip
\usepackage{amsmath}
\usepackage[english]{babel}

\def\be{\begin{equation}}
\def\ee{\end{equation}}

\def\A{\mathbb A}
\def\B{\mathbb B}
\def\Z{\mathbb Z}
\def\N{\mathbb N}
\def\R{\mathbb R}

\usepackage{todonotes}

\begin{document}
%%%%%%%%%%%%%%%%%%%%%%%%%%%%%

\title{
{\bf Klein four-group and Darboux duality\\
 in conformal mechanics  
 }}

\author{{\bf  Luis Inzunza and Mikhail S. Plyushchay} 
 \\
[8pt]
{\small \textit{Departamento de F\'{\i}sica,
Universidad de Santiago de Chile, Casilla 307, Santiago,
Chile  }}\\
[4pt]
 \sl{\small{E-mails:   
\textcolor{blue}{luis.inzunza@usach.cl},
\textcolor{blue}{mikhail.plyushchay@usach.cl}
}}
}
\date{}
\maketitle

\begin{abstract}
We study the Klein four-group $(K_4)$
 symmetry of the time-dependent Schr\"odinger equation
 for  the conformal mechanics model
of de Alfaro-Fubini-Furlan (AFF) with confining harmonic potential 
 and coupling constant 
$g=\nu(\nu+1)\geq -1/4$.
We show that it undergoes a 
 complete or partial (at  half-integer $\nu$)  breaking  on eigenstates 
 of the system,
 and is 
 the automorphism of the  $\mathfrak{osp}(2,2)$
  superconformal  symmetry 
 in super-extensions 
 of the model by
inducing a transformation between the exact and spontaneously 
broken phases of $\mathcal{N}=2$  
Poincar\'e supersymmetry.
 We exploit the $K_4$  symmetry and its  
 relation with  the conformal 
 symmetry 
 to construct   the dual Darboux transformations
 which generate spectrally shifted pairs of 
 the rationally deformed AFF models.
Two distinct pairs of intertwining operators 
originated from Darboux duality allow us 
to construct complete sets of the spectrum generating 
ladder operators 
that identify specific  finite-gap structure 
of a  deformed system 
and generate three distinct  related versions
of  nonlinearly deformed  
$\mathfrak{sl}(2,\R)$
algebra  as its symmetry.
We show that at  half-integer  $\nu$, the Jordan states 
associated with confluent Darboux transformations 
enter the construction,
and  the spectrum
of rationally deformed AFF systems undergoes structural changes.
\end{abstract}

\vskip.5cm\noindent

\section{Introduction}
In quantum mechanics, symmetries  map the states of a system
into  its states. If the ground state is invariant under the corresponding transformations,
one says that the symmetry is unbroken, otherwise symmetry is (spontaneously) broken. 
Sometimes,  along with a continuous group of symmetry transformations, 
a  discrete symmetry group appears \cite{Symmetries},
and a nontrivial interplay may occur between  both types of symmetries.
An interesting and important case from this point of view 
 is presented by the conformal mechanics model
of de Alfaro, Fubini and Furlan (AFF) \cite{AFF}
with confining harmonic potential and coupling constant 
$g=\nu(\nu+1)\geq-1/4$~\footnote{The AFF model 
is  a two-particle Calogero system  \cite{Calogero1} 
with included confining harmonic potential term
but omitted center of mass  coordinate. 
Its  Schr\"odinger operator also is known as 
the Gol'dman-Krivchenkov Hamiltonian \cite{GoldKri}.}.
Its  non-relativistic conformal symmetry 
and supersymmetric extensions \cite{SCM1,SCM2,SCM3,SCM4,SCM5,SCM6}
find a variety of  interesting applications including  the particles dynamics
 in black hole backgrounds 
 \cite{BlackHold1,BlackHold2,BlackHold3,BlackHold4,BlackHold5,Galaj},
 cosmology \cite{DGH,PioWal,CGGH},
 non-relativistic  AdS/CFT correspondence \cite{GAdS1,GAdS2,BarFue,Jack,Oka1}, QCD confinement problem
 \cite{App1,Brod2}, and physics of Bose-Einstein condensates  \cite{App2,App3}.
\vskip0.1cm

 On the other hand, the time-dependent Schr\"odinger equation for the AFF conformal
 mechanics model reveals  a discrete 
 Klein four-group  symmetry 
 generated by transformation of the parameter 
 $\nu\rightarrow -\nu-1$,
 and by the spatial Wick rotation 
$x\rightarrow ix$ accompanied by the time reflection $t\rightarrow -t$.
In the picture of the stationary Schr\"odinger equation
the time reflection transforms into 
the change of the eigenvalue's sign  $E\rightarrow -E$. 
The discrete symmetry $K_4$, however,  turns out to be 
completely broken at the level of the quantum states 
when $\nu$ is not a half-integer number\,: application 
 of the group generators to physical eigenstates produces 
formal eigenstates which do not satisfy the necessary 
boundary conditions. In the case of half-integer values of the parameter $\nu$
the $K_4$ discrete symmetry breaks partially, and 
transformation $\nu\rightarrow -\nu-1$, as we shall see,
 turns out to be a  
true symmetry  nontrivially 
realized
on the spectrum of the system.  
In  the physics of anyons, where 
the AFF model is used to generate the transmutation  of statistics,
half-integer values of $\nu$ correspond  
 to the two-particle system of  identical fermions
  \cite{LeiMyr,MKWil,Poly}. 
 In the context of the problem we consider  here,
 even though the new solutions with arbitrary value of $\nu$ 
 generated by transformations of the discrete group are not acceptable from 
the physical point of view, the analogs of such non-physical 
states in other quantum systems 
are  used  to produce 
new solvable potentials and supersymmetric extensions via the 
(generalized) Darboux transformations
 \cite{Darboux,Crum,MatSal,Witten,Cooper,Krein,Adler}.
 They also are used, for example,  for the construction of multi-soliton and finite-gap 
 solutions to the  Korteweg-de Vries equation \cite{MatSal,AMGP,AranPly}.
 In particular, solutions with  behavior typical for extreme waves are constructed
 in \cite{JM} based on non-physical states of the AFF model without the 
 harmonic trap.

The implications of the presence of the $K_4$  discrete symmetry group in the AFF model, 
and its  interplay with conformal and  superconformal symmetries,
surprisingly,  
have not been investigated yet in the literature. 
 
In this work, we study in detail the action of transformations 
of the Klein four-group on the states of the AFF system,
its relation to the  conformal symmetry,
and its nontrivial role in $\mathcal{N}=2$ super-extensions  of the AFF model
and their $\mathfrak{osp}(2,2)$  superconformal symmetry.
This superconformal symmetry, as will be shown, is based essentially 
on the simplest case of the Darboux dual schemes which produce
the same but  spectrally shifted pairs of the quantum systems.
Based on this observation,
with the help of the Klein four-group transformations we address the problem of 
 construction of the general Darboux dual schemes  to generate 
 infinite families of the new exactly solvable systems given by 
rational deformations of the conformal mechanics systems 
with arbitrary values of the parameter $\nu$.
The general dual Darboux schemes that we obtain produce 
the same but  spectrally shifted pairs of rationally deformed AFF models.
As a consequence, their distinct intertwining operators 
allow us  to construct  the complete sets of the spectrum  generating
ladder operators for them and identify  the 
nonlinearly deformed versions  of conformal 
$\mathfrak{sl}(2,\R)$ algebra  
which describe their symmetries.
In this way  we generalize our  earlier results 
obtained for the restricted case of the AFF model 
with integer values of $\nu$ only
\cite{CIP,LM2},  that were based 
on the Darboux transformations of the quantum
harmonic oscillator.
Coherently with the indicated above peculiarity  
of the half-integer values of the parameter $\nu$
from the point of view of the Klein four-group 
transformations, we will see
how the Jordan states \cite{Confluent3,confDarb,Confluent4,Confluent1,CarPly,JM,Confluent2}
 enter the construction 
at $\nu=\Z+1/2$ 
via the confluent Darboux transformations.
We also trace out the structural changes in the spectra of the rationally deformed 
AFF systems which happen at half-integer 
$\nu$ under continuous variation of this parameter.
\vskip0.1cm

The remainder of the paper is organized as follows.
In Section \ref{Dar-transform}, we summarize the basic ingredients of 
the Darboux transformations, and in their context, consider  the Jordan states. 
In Section~\ref{AFF-model}, we briefly present the AFF model together 
with its conformal symmetry and solutions of equations of motion 
provided by it, and discuss shortly the  introduction of the scale which turns out to be related to 
the holographic  QCD,
the AdS${}_2$ isometry and the AdS/CFT correspondence,
 and Dirac's different forms of relativistic dynamics. 
Section \ref{AFFsym} is devoted to the discussion of the
discrete Klein four-group symmetry of the time-dependent 
Schr\"odinger equation for the AFF model and its action
on physical states, its relation to the 
conformal symmetry, and the  role played by the discrete $K_4$ 
group  in the $\mathcal{N}=2$ super-extensions of the 
AFF system generated by  the 
simplest  dual Darboux  schemes.
The algorithm of construction of  the general  dual Darboux schemes is 
 developed in Section \ref{Mirror}.   
In Section  \ref{Ladders}  we  list some Darboux schemes which produce 
distinct infinite families of new 
exactly solvable systems described  by rational deformations
of the AFF model with arbitrary number of gaps 
introduced into the equidistant spectrum of the original system.
For each such a Darboux scheme we construct its dual
that plays a key role in identifying the spectral 
properties of the generated system  and its symmetry.
We also  trace out there the structural changes in the spectra
of the obtained systems under continuous variation 
of the parameter $\nu$.
In Section  \ref{SecIntLad}, we use the intertwining operators 
of the dual Darboux schemes to construct the complete sets
of the spectrum generating ladder operators  of the rationally
deformed AFF systems. These higher order differential 
operators detect and describe  the `finite-gap' structure of the
 spectra of the corresponding systems, 
and generate three distinct but related versions of  the nonlinearly
deformed  $\mathfrak{sl}(2,\R)$ conformal 
algebra  which describe their symmetries.
 The application of the general results of Sections 
 \ref{Mirror}, \ref{Ladders} and  \ref{SecIntLad}
 is  illustrated by an example presented in  Section  \ref{Examples}.
In Section \ref{Discussion} we summarize  the results and 
discuss some problems to be interesting for further investigation.  
In three Appendices some  technical details necessary for the main text 
are presented.

\section{Generalized Darboux transformations}
\label{Dar-transform}
In this section we summarize some properties 
of the generalized Darboux transformations 
which will be employed in what follows.

\subsection{Darboux transformations and intertwining operators}
Consider the equation
\be
\label{Sch}
L_0\psi_\lambda=\lambda\psi_\lambda\,,\qquad
L_0=
-\frac{d^2}{dx^2}+V(x)\,,
\ee
corresponding to the eigenvalue problem of 
a Schr\"odinger type operator $L_0$. 
In this section 
we treat Eq. (\ref{Sch}) as a formal second order differential equation
on some interval $(a,b)$,
and in the following sections 
we take care about its physical nature.
Suppose we have a set of solutions $\psi_{k}$
corresponding to eigenvalues $\lambda_{k}$,
$k=1,\ldots,n$. We use them as seed states 
for generalized Darboux transformation and  generate the associated  eigenvalue problem 
\be
\label{Dar}
L_{[n]}\Psi_{\lambda}=
\lambda\Psi_\lambda\,,\qquad
L_{[n]}=
-\frac{d^2}{dx^2}+V(x)-2\frac{d^2}{dx^2}
\ln W(\psi_{1},\ldots,\psi_{n})\,.
\ee
If the set of the seed states is chosen in such a way 
that the Wronskian $W$ takes non-zero values on $(a,b)$,
then potential of the generated system will also be nonsingular 
there.
 In general case,  
solutions of (\ref{Dar}) are generated 
from corresponding solutions of Eq. (\ref{Sch}),
\begin{equation}
\label{Darstates}
\Psi_{\lambda}=\frac{W(\psi_{1},\ldots,\psi_{n},\psi_{\lambda})}{W(\psi_{1},\ldots,\psi_{n})}=\A_{n}\psi_{\lambda}\,,
\end{equation}
where $\A_{n}$ is the  differential operator of
order $n$  defined recursively as 
\be
\label{generic-inter}
\A_{n}=A_n\ldots A_1\,,\qquad A_k=\A_{k-1}\psi_k\frac{d}{dx}\left(\frac{1}{\A_{k-1}\psi_k}\right),
\qquad k=1,\ldots,n,\qquad \A_0=1\,.
\ee
By the construction,  $\ker\A_n=\text{span}\{\psi_{1},\ldots,\psi_{n}\}$.
Operator $\A_n$ and its Hermitian conjugate $\A_n^\dagger$ intertwine the operators 
$L_0$ and $L_{[n]}$,
\be
\label{inter-gen}
\A_nL_0=L_{[n]}\A_n\,,\qquad \A_n^\dagger L_{[n]}=L_0\A_n^\dagger\,,
\ee
and satisfy relations 
\be
\label{poly1}
\A_n^\dagger\A_n=\prod_{k=1}^{n}(L_{0}-\lambda_k)\,,\qquad
\A_n\A_n^\dagger=\prod_{k=1}^{n}(L_{[n]}-\lambda_k)\,.
\ee
From the first equation in (\ref{poly1}) one can find that
$\ker\A_n^\dagger=\text{span}\{\A_n\widetilde{\psi}_{1},\ldots,\A_n\widetilde{\psi}_{n}\}$,
where 
\begin{equation}\label{tildepsi}
\widetilde{\psi}_{\lambda}=\psi_{\lambda} \int^x\frac{d\zeta}{(\psi_{\lambda}(\zeta))^2}\,
\end{equation}
is a linear independent solution of Eq. (\ref{Sch})
with  the same eigenvalue $\lambda$,
$W(\psi_\lambda,  \widetilde{\psi}_{\lambda})=1$.
Similarly to (\ref{Darstates}), $\A_{n}^\dagger \Psi_{\lambda}=\psi_{\lambda}$
for  $\Psi_\lambda\notin \text{ker}\,\A_{n}^\dagger$,
and 
$$
\A_{n}^\dagger \widetilde{(\A_n\widetilde{\psi}_k)}=\psi_k\in \text{ker}\,\A_n\,.
$$
Here and in what follows we consider equalities 
between wave functions and Wronskians  
in  `up to a multiplicative constant' sense
when the corresponding constant will be inessential.

The generalized 
Darboux transformation possesses the iterative property according to 
which system (\ref{Dar})  
can be generated alternatively  via  successive 
Darboux transformations\footnote{Intermediate systems
appearing in such a way may have a singular nature on the interval $(a,b)$  \cite{AMGP,PlySch,CarPly2}.}. 
This property allows us to get some useful Wronskian identities.   
Suppose that we have two collections of (formal) eigenstates
 of (\ref{Sch}), $\{\phi_n\}=(\phi_1,\ldots,\phi_n)$ and 
$\{\varphi_l\}=(\varphi_1,\ldots,\varphi_l)$. 
In the first step, we generate a Darboux transformation
by taking the first collection as the set of the 
seed states, and obtain 
 the  intermediate Hamiltonian operator with potential 
$V_1=V(x)-2(\ln W(\{\phi_n\}) )''$.
In this way,  the states of the second collection $\{\varphi_l\}$ 
will be mapped into  the set of (formal in general case) eigenstates 
$\{\A_n\varphi_l\}=(\A_n\varphi_1,\ldots,\A_n\varphi_l)$
of the intermediate system,
where $\A_n$ is the 
corresponding intertwining operator of order $n$ constructed following (\ref{generic-inter}). 
Then, employing these states as the seed states 
for  a second Darboux transformation, 
we finally obtain a Schr\"odinger  operator with a potential 
$V_2=V_1(x)-2(\ln W(\{\A_n\varphi_l\}))''$.
Having in mind that  
the same result will be produced by a one-step generalized 
Darboux transformation 
based on the whole set of the chosen eigenstates of the system $L_0$,
we obtain  the equality 
\begin{equation}
\label{id1}
W(\{\phi_n\})W(\{\A_n\varphi_l\})=W(\phi_1,\ldots,\phi_n,\varphi_1,\ldots,\varphi_l)\,.
\end{equation}

Consider now the set of two states corresponding to a same eigenvalue 
${\lambda_j}$,   $\{\phi_{2}\}=(\phi_{1}=\psi_{j},\phi_{2}=\widetilde{\psi}_{j})$.  
In this case $W(\psi_{j},\widetilde{\psi}_{j})=1$,  and the corresponding intertwining operator 
reduces to  $\A_2=-(L_0-\lambda_j)$.
Using this observation and  Eq. (\ref{id1}), we derive the equality
$W(\psi_j,\widetilde{\psi}_j,\varphi_1,\ldots,\varphi_l)=W(\{\varphi_l\})$, 
which is generalized for the relation
\be
\label{ide2}
W(\psi_{1},\widetilde{\psi}_{1},\ldots,\psi_{s},\widetilde{\psi}_{s},\varphi_1,\ldots,\varphi_l)=W(\{\varphi_l\})\,.
\ee
In the case when  functions  $\varphi_1,\ldots,\varphi_l$ are not obligatorily to be eigenstates 
of  the operator $L_0$, the 
last relation changes  for
\begin{eqnarray}
\label{ide2+}
&W(\psi_{1},\widetilde{\psi}_{1},\ldots,\psi_{s},\widetilde{\psi}_{s},\varphi_1,\ldots,\varphi_l)=
W\left(\{\prod_{k=1}^s(-L_0+\lambda_k)\varphi_l\}\right).&
\end{eqnarray}

\subsection{Jordan states}
For a given Schr\"odinger operator $L_0$, one
 can construct a certain set of functions which are not its eigentstates
but are annihilated by the action of a certain polynomial 
of $L_0$. Functions of such a nature can be related with the confluent 
Darboux transformations and 
are  identified as Jordan states \cite{confDarb}.
They  were used, for 
example, in the construction 
of  isospectral  deformations of the harmonic oscillator systems \cite{CarPly,InzPly},
and also they appeared in the context of solutions 
to the KdV equation \cite{JM}. In this subsection we construct Jordan states that are solutions of 
the fourth order differential equation 
$(L_0-\lambda_*)^2\chi_*=0$. 
They will play an important role in subsequent consideration.

We employ the following 
approach:  take an  eigenstate $\psi_*$ corresponding to eigenvalue $\lambda_*$ 
as a seed state of the  Darboux transformation. This provides us with the
first order differential  operators    
\be
A_{\psi_*}=\psi_*\frac{d}{dx}\left(\frac{1}{\psi_*}\right)\,,\qquad
A_{\psi_*}^\dagger=-\frac{1}{\psi_*}\frac{d}{dx}\psi_*\,.
\ee  
According to Eq.  (\ref{poly1}),  their product gives us the shifted Schr\"odinger operator 
$A_{\psi_*}^\dagger A_{\psi_*}=L_0-\lambda_*$, whose kernel is spanned by
the linear independent states $\psi_*$ and $\widetilde{\psi}_*$. 
The problem of constructing Jordan states reduces then  to solving 
equations 
\be\label{Omega12}
A_{\psi_*}^\dagger A_{\psi_*}\Omega_*=(L_0-\lambda_*)\Omega_*=\psi_*\,,\qquad
A_{\psi_*}^\dagger A_{\psi_*}\breve{\Omega}_*=(L_0-\lambda_*)\breve{\Omega}_*=\widetilde{\psi}_*\,.
\ee
Their solutions are given, up to a linear combination of
 $\psi_*$ and  $\widetilde{\psi}_*$,  by particular solutions of respective
  inhomogeneous 
 equations, 
\begin{eqnarray}
\label{omega1}
\Omega_*=\psi_*\int_{a}^{x}\frac{d\zeta}{\psi_*^2(\zeta)}\int_{\zeta}^{b}\psi_*^2(\eta)d\eta
\,,\qquad
\breve{\Omega}_*=\psi_*\int_{a}^{x}\frac{d\zeta}{\psi_*^2(\zeta)}\int_{\zeta}^{b}\psi_*
(\eta)\widetilde{\psi}_*(\eta)d\eta\,.
\end{eqnarray}
Here the  integration limits  are chosen coherently 
with  the region where  the operator $L_0$  is defined,
and we have the relations
\begin{eqnarray}
\label{JorWronskian}
W(\psi_*,\Omega_*)=\int_{x}^{b}\psi_{*}^2d\zeta\,,\qquad
W(\psi_*,\breve{\Omega}_*)=\int_{x}^{b}\psi_{*}\widetilde{\psi}_*d\zeta\,,
\end{eqnarray}
which will be useful to produce nonsingular confluent 
Darboux transformations.

Let us  inspect  now the role of Jordan states (\ref{omega1})  in  Darboux transformations
generated by a set of the seed states $\{\psi_n\}$.     
The intertwining operator (\ref{generic-inter}) and 
equations (\ref{inter-gen}) and (\ref{Omega12}) give us the relations
\be
\label{Dar-jor}
\A_n\psi_{*}=(L_{[n]}-\lambda_{*})\A_n\Omega_{*}\,,\qquad
\A_n\widetilde{\psi}_{*}=(L_{[n]}-\lambda_{*})\A_n\breve{\Omega}_{*}\,.
\ee
If the state $\psi_*$ (or $\widetilde{\psi}_*$) is annihilated by 
$\A_n$, i.e. if the set of the seed states $\{\psi_n\}$ includes  $\psi_*$ (or $\widetilde{\psi}_*$),
the function  $\A_n\Omega_{*}$ (or $\A_n\breve{\Omega}_{*}$) 
will be an eigenstate of $L_{[n]}$ with eigenvalue $\lambda_*$
which is available to produce another Darboux transformation 
if we consider  $L_{[n]}$ as an intermediate system.   
Otherwise,  
the indicated  function is a Jordan state of 
$L_{[n]}$, and in correspondence  with (\ref{omega1}) we have 
\begin{eqnarray}
\label{omega2}
&\A_n\Omega_*=(\A_n\psi_*)\int_{a}^{x}\frac{d\zeta}{(\A_n\psi_*)^2(\zeta)}\int_{\zeta}^{b}(\A_n\psi_*)^2(\eta)d\eta
\,,\qquad&\\&
\A_n\breve{\Omega}_*=(\A_n\psi_*)\int_{a}^{x}\frac{d\zeta}{(\A_n\psi_*)^2(\zeta)}\int_{\zeta}^{b}(\A_n\psi_*)
(\eta)\widetilde{\A_n\psi_*}(\eta)d\eta\,&
\end{eqnarray}
up to a linear combination with $\A_n\psi_*$ and  $\widetilde{\A_n\psi_*}$.

Having in mind that Jordan states appear 
naturally  in the confluent generalized Darboux transformations \cite{confDarb}, 
one can consider directly a generalized Darboux transformation 
based on the  following set of the seed states\,:
 $(\psi_1,\Omega_1,\ldots,\psi_n,\Omega_n)$.
 This generates a Darboux-transformed  
 system which we denote by 
 $\widehat{L}_{[2n]}$. The  intertwining operator 
$\A_{2n}^{\Omega}$ as a differential operator of order $2n$ is built 
according to the same rule (\ref{generic-inter}),
but with the inclusion  of Jordan states into the set
of generating functions. 
By the construction, this operator annihilates the chosen $2n$ seed  states,
and  one can show that 
\be
\label{Polly2}
(\A_{2n}^\Omega)^\dagger\A_{2n}^\Omega=\prod_{i=1}^{n}(L-\lambda_i)^2\,,\qquad
\A_{2n}^\Omega(\A_{2n}^\Omega)^\dagger=\prod_{i}^{n}(\widehat{L}_{[2n]}-\lambda_i)^2\,.
\ee
This, in particular, means that 
$\ker(\A_{2n}^\Omega)^\dagger=\text{span}\{\A_{2n}^\Omega\widetilde{\psi}_{1},
\A_{2n}^\Omega\breve{\Omega}_{1},\ldots,\A_{2n}^{\Omega}\widetilde{\psi}_{n},\A_{2n}^{\Omega}\breve{\Omega}_{n}\}$.
In the context of generalized Darboux transformations based 
on a mixture of 
eigenstates and Jordan states,  
a useful relation
\begin{equation}
\label{tech1}
W(\psi_{*},\widetilde{\psi}_{*},\Omega_*,\breve{\Omega}_*,\varphi_1,\ldots,\varphi_l)=
W(\varphi_1,\ldots,\varphi_l)\, 
\end{equation}
can be obtained by employing Eq. (\ref{ide2+})
with $s=1$, and  Eqs. (\ref{Omega12}) and (\ref{ide2}),
Here  we  imply that
$\varphi_i$ with $i=1,\ldots,l$ is the set of solutions of equation (\ref{Sch})
with $\lambda_i\neq \lambda_*$.

%%%%%%%%%%%%%%%%%%%%%%%%%%%%
\section{AFF conformal mechanics model}
\label{AFF-model}
%%%%%%%%%%%%%%%%%%%%%%%%%%%%

Consider now the one-dimensional system given by the action 
\cite{AFF} 
\be
\label{conformalaction}
I[q]=\int L(q,\dot{q}) dt\,, \quad
L=\frac{1}{2}\left(\dot{q}^2-\frac{g}{q^2}\right)\,, 
\ee
where $q>0$ has dimension $[q]=[\sqrt{t}]$, 
and $g$ is a dimensionless coupling constant 
which classically is assumed to be positive 
to avoid the `problem of fall to  the center'. 
 System (\ref{conformalaction}) is characterized 
 by conformal symmetry $Conf (\R^1)$
 that canonically is generated by the
 Hamiltonian $H_g$, 
  the dilatation 
 generator $D$, and generator of special conformal transformations  
 $K$,
 \begin{equation}
 \label{conformalgenerators}
 H_g=\frac{1}{2}(p^2+\frac{g}{q^2})\,,\qquad
 D=\frac{1}{4}(qp+pq) -H_gt\,,\qquad
 K=\frac{1}{2}q^2-2Dt-H_gt^2\,,
\end{equation}      
 where $p=\dot{q}$. 
 These are the integrals of motion that satisfy the equation 
 of the form
 $\frac{d}{dt}{A}=\frac{\partial A}{\partial t} + \{A,H\}=0$.
 They  obey  the 
 $\mathfrak{so}(2,1)$ algebra\footnote{To treat  integrals
 (\ref{conformalgenerators}) as generators of the Lorentz algebra 
 $\mathfrak{so}(2,1)$ requires an  introduction of a constant 
 $\sigma$ with dimension of time  (squared length)  to make 
 $H_g$ and $K$ dimensionless,
  $H_g\rightarrow \sigma H_g$, $K\rightarrow \sigma^{-1} K$,
  that, however, does not change the form of algebra (\ref{so(2,1)}).}
\begin{equation}
\label{so(2,1)}
\{D,H_g\}=H_g\,,\qquad
\{D,K\}=-K\,,\qquad
\{H_g,K\}=-2D\,,
\end{equation}
which is isomorphic to the algebra $\mathfrak{sl}(2,\R)$ 
of the conformal symmetry $Conf (\R^1)$.
Classical algebra (\ref{so(2,1)}) is characterized by the 
Casimir invariant
$Q=KH_g-D^2$ that takes the   value $Q=\frac{1}{4}g$.
Last relation from (\ref{conformalgenerators}) 
gives us  solution to the equation of motion for the system
(\ref{conformalaction}), 
\be\label{q(t)}
q(t)=\sqrt{2(at^2+2bt+c)}\,,
\ee
where real-valued  constants $a$, $b$ and $c$
 correspond to the  
values of the integrals  $H_g$, $D$ and $K$, respectively,
and thus are restricted by conditions $a>0$, $c>0$, and relation  
$ac-b^2=g/4$ that guarantees positive values for the
square root argument.

At the quantum level the non-commutativity 
of $D$ and $K$ with Hamiltonian $H_g$ 
in correspondence with the Poisson bracket relations
(\ref{so(2,1)}) means that 
they do not generate symmetries in the usual sense of relating degenerate states,
but rather they can be used to
 relate the states with different eigenvalues of 
the Hamiltonian operator.

Conformal symmetry  (\ref{so(2,1)})  is a dynamical 
symmetry for the system (\ref{conformalaction}),
but it also is the isometry of AdS${}_2$. This underlies
the interest to the model \cite{AFF}  in the context of the AdS/CFT 
correspondence and its  diverse applications, in particular,  
related to the appearance of scale in nominally conformal theories \cite{App1}. 

Original system (\ref{conformalaction}) has no scale,
but the scale  emerges in the theory via  
a mechanism  described by  de Alfaro, Fubini and Furlan \cite{AFF}. 
This happens as follows.
First relation in (\ref{so(2,1)}) 
and second relation in (\ref{conformalgenerators}) reflect the fact that action 
(\ref{conformalaction}) is explicitly invariant under 
the scale transformations $q\rightarrow e^\alpha q$,
  $t\rightarrow e^{2\alpha} t$, $\alpha\in\R$, in accordance with which 
 at the quantum level the dilatation operator
 generates the  
 transformation 
 $H_g\rightarrow e^{-2\alpha}H_g$.
Consequently,  
 the Hamiltonian operator of the quantum version 
 of the system (\ref{conformalaction})
 has a continuous spectrum in which  there is no conformal invariant 
 ground state.
  This  can be  related with the nature of the evolution coordinate $t$, 
  which is not a good global coordinate on AdS${}_2$ \cite{BlackHold2}. 
  To resolve this problem, one can  consider the following 
  change of the variables 
  \begin{equation}
  \label{trans}
  y(t)=\frac{q(t)}{\sqrt{u+vt+wt^2}}\,, \qquad d\tau=\frac{dt}{u+vt+wt^2}\,,
  \end{equation}
where $u>0$, $v$ and $w>0$ are real
constants with dimensions $[u]=1$, $[v]=1/t$ and $[w]=1/t^2$. By replacement (\ref{trans}),
action 
 (\ref{conformalaction}) transforms into 
 \begin{eqnarray}
\label{conformalaction2}
&\int\mathcal{L}(y,y')d\tau
+
\frac{1}{4}\int d\tau\frac{d}{d\tau}[(v+2wt(\tau))q^2(t(\tau))]=
I[y]+I_{surface}\,,\,\,&
\end{eqnarray} 
 where $\mathcal{L}(y,y')=\frac{1}{2}(y'^2-\frac{g}{y^2}-\omega^2y^2)$,
 $y'=\frac{dy}{d\tau}$, and $\omega^2=(4wu-v^2)/4$. 
 {}From the action $I[y]=\int \mathcal{L}d\tau$, we obtain the new time translation generator  
\begin{eqnarray}
\label{mostgeneralH}
\mathcal{H}_g=\frac{1}{2}\left(p^2+\frac{g}{y^2}+\omega^2 y^2\right)\,,\qquad p=y'\,,
\end{eqnarray}
which is a compact $\mathfrak{sl}(2,\R)$ generator 
when $\omega^2=(4wu-v^2)/4>0$,
whose quantum analog, like quantum analog of $H_g$,
 has a spectrum restricted from below
 when $g\geq-1/4$ \cite{LLQM,KirLoy}. The new evolution 
parameter $\tau=\frac{1}{\omega}\arctan(\frac{v+2wt}{2\omega})$ 
varies in the 
finite interval 
$(-\frac{\pi}{2\omega},\frac{\pi}{2\omega})$,
and new Hamiltonian (\ref{mostgeneralH}) is conjugate to this 
good global time coordinate.
In the  context of black hole physics the AFF suggestion simply
amounts to an improved choice of time coordinate \cite{BlackHold2,BlackHold5}.
As $\omega$  is a dimensionfull  parameter, $[\omega]=[1/t]$,
(\ref{mostgeneralH}) breaks the manifest scale invariance
of the original system  (\ref{conformalaction}), 
and via such a basic mechanism the mass and length scales 
are introduced in holographic QCD (often referred to as ``AdS/QCD")
\cite{App1,Brod2}.

In spite of the introduced scale, 
 the action of the new system is conformal invariant 
as we will see now.
The dilatation generator $\mathcal{D}$ and 
the conformal transformation generator $\mathcal{K}$ associated with the 
action 
$I[y]$ are given by the
explicitly depending on time $\tau$ integrals 
\begin{eqnarray}
\label{NHgenD}
&\mathcal{D}=\frac{1}{2}\left(yp\cos(2\omega \tau)+\left(2\omega y^2-
\mathcal{H}_g{\omega}^{-1}\right)\sin(2\omega \tau)\right)\,,&\\
&\mathcal{K}= \frac{1}{2}\left(y^2 \cos(2\omega \tau) -yp{\omega}^{-1}\sin(2\omega \tau)-\mathcal{H}_g
\omega^{-2}\left(\cos(2\omega \tau) -1\right)\right)\,.\label{NHgenH}&
\end{eqnarray}
Via the Noether theorem, integrals (\ref{NHgenD}) and (\ref{NHgenH})
are related to the following infinitesimal symmetries of the action:
\begin{eqnarray}\label{deltaD}
&\delta y=\epsilon_{\mathcal{D}}y\cos(2\omega \tau)\,,
\quad
\delta \tau=\epsilon_{\mathcal{D}}{\omega}^{-1}\sin(2\omega \tau)\,,
\,\,
\delta(\mathcal{L}d\tau)=\frac{d}{d\tau}\left(
-\epsilon_{\mathcal{D}}\omega y^2\sin(2\omega t)\right)d\tau\,,&\\
\label{deltaK}
&\delta y=-\epsilon_{\mathcal{K}}\omega^{-1}\sin(2\omega \tau)\,,
\,\,
\delta \tau=\epsilon_{\mathcal{K}}\omega^{-2}(\cos(2\omega \tau)-1)\,,
\,\,
\delta(\mathcal{L}d\tau)=\frac{d}{d\tau}\left(
-\epsilon_{\mathcal{K}}\omega y^2\cos(2\omega t)\right)d\tau\,.\qquad&
\end{eqnarray}
Integrals $\mathcal{H}_g$,  $\mathcal{D}$ and $\mathcal{K}$
of the `regularized' AFF system 
generate  the Newton-Hooke symmetry \cite{GalajNH,NHSymmetry,SCM6}
\begin{equation}\label{NHalg}
\{\mathcal{H}_g,\mathcal{D}\}=-(\mathcal{H}_g-2\omega^2 \mathcal{K})\,,\qquad
\{\mathcal{H}_g,\mathcal{K}\}=-2\mathcal{D}\,,\qquad
\{\mathcal{D},\mathcal{K}\}=-\mathcal{K}\,,
\end{equation}
whose Casimir invariant is $\mathcal{Q}=\mathcal{K}\mathcal{H}_g-\mathcal{D}^2-\omega^2\mathcal{K}^2=g/4$.
Using Eqs. (\ref{NHgenD}) and (\ref{NHgenH}),
one can find solution to the equation of motion of the 
system (\ref{conformalaction2}),
\be\label{ytau}
y(\tau)=\omega^{-1}\sqrt{\frac{2}{1+\sin^2(2\omega \tau)}}
\sqrt{a\sin^2(\omega \tau)+b\omega\sin(2\omega \tau)+
c\omega^2\cos(2\omega \tau)}\,,
\ee
where $a>0$, $b$ and $c>0$ are constants 
corresponding to the values of the integrals
 $\mathcal{H}_g$, $\mathcal{D}$ and $\mathcal{K}$, respectively, and  
obeying the relation 
$ac-b^2-\omega^2c^2=g/4$. {}From the explicit form of solution
we see that it is periodic with the period $T=\pi/\omega$ not depending on the value 
of the coupling constant\footnote{System
given by Hamiltonian (\ref{mostgeneralH}) is an isoperiodic deformation 
of the half-harmonic oscillator of frequency $\omega$ \cite{Aso}.} $g$.
The finite interval in which the  evolution parameter $\tau$
varies  corresponds to the period of the motion of the 
system (\ref{conformalaction2}), and one can consider
$\tau$ as the compact evolution parameter that
takes values on the closed interval 
$[-\frac{\pi}{2\omega},\frac{\pi}{2\omega}]$
with identified ends.

In the limit $\omega\rightarrow 0$, Hamiltonian $\mathcal{H}_g$
reduces to the Hamiltonian $H_g$ of the model (\ref{conformalaction}),
the integrals $\mathcal{D}$ and $\mathcal{K}$ reduce to the integrals
$D$ and $K$,  the algebra 
(\ref{NHalg})  takes the form (\ref{so(2,1)}), and bounded periodic 
solution (\ref{ytau})
transforms into solution (\ref{q(t)}) for the system  
(\ref{conformalaction}), which describes the unbounded 
motion of a particle in potential $g/q^2$.
The  symmetry transformations of the system (\ref{conformalaction})
generated by $D$ and $K$ are recovered from 
(\ref{deltaD}) and (\ref{deltaK}). 
In particular, the nontrivial total time derivative in the last relation 
in (\ref{deltaD}) transforms in this limit into $\delta(Ldt)=0$.
This reflects the difference between manifest scale symmetry of the system
(\ref{conformalaction}) and
more complicated, `hidden' form of the dilatation symmetry of the 
system given by Lagrangian $\mathcal{L}$.
In application of  the system (\ref{conformalaction2})
to the problem of confinement in QCD,
parameter $\omega$ introduces a mass scale, 
while the evolution parameter  $\tau$ 
finds an interpretation as  the difference of
light-front times between events involving quarks and antiquarks
in mesons  \cite{App1}.

To clarify further the relation between systems 
(\ref{conformalaction}) and 
(\ref{conformalaction2}), 
we note that 
at  $\tau=t=0$
we have $y(0)=q(0)/\sqrt{u}$, $y'(0)=\sqrt{u}\dot{q}(0)-\frac{vq(0)}{2\sqrt{u}}$, and therefore the Hamiltonian 
(\ref{mostgeneralH}) becomes   
 \begin{equation}
\label{mostgeneralH2}
\mathcal{H}_g=uH_g-vD+wK\,.
\end{equation}
This shows that transformation (\ref{trans}) 
means in fact that a certain  linear combination of 
 $\mathfrak{sl}(2,\R)$ 
generators 
is used as a new generator of time translations.
To be more specific, 
we note that
the  $\mathfrak{sl}(2,\R)$  algebra 
can be presented in the form 
of   $\mathfrak{so}(2,1)$ Lorentz algebra
\be\label{21J}
\{J^\mu,J^\nu\}=\epsilon^{\mu\nu\lambda}J_\lambda,
\ee
where $\mu,\nu,\lambda=0,1,2,$
$J_\mu=\eta_{\mu\nu}J^\nu$, $\eta_{\mu\nu}=\text{diag}\,(-1,1,1)$,
and $\epsilon^{\mu\nu\lambda}$ is the Levi-Civita totally antisymmetric tensor, 
$\epsilon^{012}=1$.
Comparison of (\ref{so(2,1)}) with (\ref{21J})
shows that the integrals $\sigma H_g$, $\sigma^{-1} K$ and $D$
of the system (\ref{conformalaction}) can be identified,  
up to $\mathfrak{so}(2,1)$ transformation,
with generators $(J^0+J^1)$, $(J^0-J^1)$ and $J^2$,
and Casimir invariant $Q$ of algebra (\ref{so(2,1)})
corresponds to the Casimir element 
$-J^\mu J_\mu$ of algebra (\ref{21J}).
Relation $-J^\mu J_\mu=g>0$
together with inequality $(J^0+J^1)>0$
describes the upper sheet of the two-sheeted hyperboloid 
in $(2+1)$ dimensional  Minkowski space with coordinates $J^1$, $J^2$, $J^0$.
According to such an identification, 
Hamiltonian $H_g$  corresponds to a non-compact 
$\mathfrak{sl}(2,\R)$ generator of parabolic type \cite{Barg,MPsl} 
being a linear combination
of generators of rotation and Lorentz boost transformations 
in $(2+1)$ dimensional  Minkowski space. 
Putting then for simplicity $u=1$, $v=0$, $w=\sigma^{-2}$,
we reduce (\ref{mostgeneralH2}) 
to the relation $\frac{1}{2}\sigma \mathcal{H}_g=J^0$
that shows that Hamiltonian 
(\ref{mostgeneralH}) is the
generator of rotations in $(2+1)$-dimensional
Minkowski space with coordinates $J^\mu$.
According to Dirac \cite{Dirac,AFN},
$H_g$ and $\mathcal{H}_g$ 
provide us with different forms 
of relativistic dynamics  on 
the upper sheet of two-sheeted hyperboloid in $(2+1)$-dimensional
Minkowski space, which by means of solutions (\ref{q(t)}) and 
(\ref{ytau}) are projected 
to configuration spaces 
with coordinates $q$ and $y$,
and are described by evolution parameters $t$ and $\tau$,
respectively.

 \vskip0.1cm
 Changing the variable $y$ for $x=\sqrt{\omega}y$, and canonically quantizing 
 (\ref{mostgeneralH}), we obtain the Hamiltonian operator  ($\hbar=1$)
\begin{eqnarray}
\label{Lg}
\hat{\mathcal{H}}_{g}=\frac{\omega}{2}\left(-\frac{d^2}{dx^2}+x^2+\frac{g}{x^2}\right)\,,
\qquad
x\in(0,\infty)\,.
\end{eqnarray}
{}From now on, we assume that  $g\geq -1/4$ 
 to ensure that the spectrum
of (\ref{Lg})  is bounded from below \cite{LLQM,KirLoy}.
In correspondence with this,
one can set in (\ref{Lg}) $g=\nu(\nu+1)$ with  $\nu\geq -1/2$.
Following \cite{Calogero1},
as a domain of  (\ref{Lg}) we take
$\{ \psi\in L^2((0,\infty),dx)\vert \psi(0^+)=0\}$.
As we will see, 
this guarantees that  $\psi'(0^+)=0$ for $\nu> 0$, while
$0<|\psi'(0^+)|<\infty$ for $\nu=0$.
We require additionally that
$\psi(x)\psi'(x)\vert_{x=0^+}=0$  for $-1/2<\nu<0$, and 
 $\psi(x)\psi'(x)\vert_{x=0^+}=c\neq \infty$
when  $\nu=-1/2$.
In all the cases the specified properties 
of wave functions will guarantee that 
probability flux vanishes at $x=0^+$.
The specified domain corresponds to  
(essentially) self-adjoint nature of the operator
(\ref{Lg})
in the case of $\nu\geq 1/2$,
and to a special case $\theta=\pi/2$ of 
one-parametric, $\theta\in [0,\pi]$, families 
of  self-adjoint extensions of $\hat{\mathcal{H}}_{g}$
in the cases of $-1/2<\nu<1/2$ and $\nu=-1/2$.
This special case is the unique value of 
$\theta$ for which the spectrum of 
(\ref{Lg}) with $-1/4\leq g<3/4$ is equidistant 
similarly to the case of $g\geq 3/4$.
For the detailed discussion of the issue of self-adjoint 
extensions of the operator (\ref{Lg})  we refer to 
 \cite{KirLoy,FPW,FalPis}.
In the next section we shall see that 
the specified  quantum Hamiltonian  operator (\ref{Lg})
corresponds, in accordance with the described above 
classical picture, to the so called discrete type 
representation $\mathcal{D}^+_\alpha$
of the  $\mathfrak{sl}(2,\R)$  algebra 
characterized by the value of the Casimir 
operator $\hat{J}^\mu \hat{J}_\mu=-\alpha(\alpha-1)$,
in which the compact generator $\hat{J}^0$
takes positive discrete values $j^0_n=n+\alpha$,
$n=0,1,\ldots$, with $\alpha=\frac{1}{2}\nu+\frac{3}{4}\geq \frac{1}{2}$
\cite{MPsl}.
\vskip0.1cm

In what follows, we identify  new, not described 
earlier in the literature,  finite discrete symmetries 
of the Schr\"odinger equation with Hamiltonian 
operator of the regularized AFF model.
We show that these finite discrete symmetries,
 described by the Klein four-group, 
are nontrivially encoded in the conformal 
symmetry of the AFF model.
We also find that this discrete symmetry has 
 peculiarities in the case of half-integer values 
 of the parameter $\nu$, and, in particular, at $\nu=-1/2$.
 The value $\nu=-1/2$, as we will see, also is special 
 in super-extensions  of the AFF model, where the transformations of 
 the Klein four-group
 appear as automorphisms of the superconformal 
 $\mathfrak{osp}(2,2)$ symmetry and relate (mutually map)
 the corresponding systems with exact and spontaneously
 broken $\mathcal{N}=2$ Poincar\'e supersymmetries.
Then we  use  these discrete  symmetries 
together with the conformal symmetry 
to generate, by means of Darboux transformations, 
infinite families of
new, exactly  solvable quantum systems
with equidistant spectra containing 
arbitrary number of gaps and described
by symmetries of the type of finite $W$ algebras, 
which represent nonlinearly deformed and extended
forms of the  conformal $\mathfrak{sl}(2,\R)$ algebra.

\section{Symmetries of the quantum  AFF model and its\\ $\mathcal{N}=2$ super-extensions}
\label{AFFsym}

In this section, we discuss some aspects of symmetry of the 
 quantum AFF model  given by the Hamiltonian operator (\ref{Lg}).
 Namely, we show that the corresponding Schr\"odinger equation
 has a finite discrete symmetry, which is broken in general case and is encoded 
 in the ``fine" structure of conformal symmetry of the system.
 It also will be shown that in the case of half-integer values of the parameter $\nu$,
 the finite discrete symmetry has some peculiarities
related  to the structure of  eigenstates and eigenvalues 
 of the quantum Hamiltonian  (\ref{Lg}). 
 These peculiarities 
 are manifested in a spectral symmetry realized in the form of a 
 ``departure from the Hilbert's hotel"  mechanism
 and in appearance of Jordan states 
 in kernels of the ladder operators 
  being non-compact
 generators of conformal symmetry.
 We also discuss here  the effect of discrete group 
 on superconformal symmetry of $\mathcal{N}=2$ 
 super-extensions  of the 
 AFF system.
 \vskip0.1cm

Consider the time-dependent Schr\"odinger equation 
of the quantum system (\ref{Lg}),
\begin{equation}
\label{timedependent1}
i\frac{\partial\Psi_\nu(x,t)}{\partial t}=
\mathcal{H}_{\nu}\Psi_\nu(x,t)\,,
\end{equation}
with
\be
\label{Lnu}
\mathcal{H}_{\nu}=-\frac{d^2}{dx^2}+x^2+\frac{\nu(\nu+1)}{x^2}\,.
\ee
Here, the operator $\mathcal{H}_\nu$ 
corresponds to  
(\ref{Lg}) with
$g(\nu)=\nu(\nu+1)$, $\nu\geq -1/4$, and 
$\omega=2$. 
Solutions of Eq. (\ref{timedependent1}) in the form
of stationary states $\Psi_\nu(x,t)=e^{-it\lambda_\nu}\psi_{\nu}(x)$
are given in terms of the well known physical eigenstates of  $\mathcal{H}_\nu$
 represented by normalized wave functions satisfying the boundary conditions
at $x=0$, 
\begin{eqnarray}
\label{physeig}
&\psi_{\nu,n}(x)=\sqrt{\frac{n!}{\Gamma(n+\nu+3/2)}}x^{\nu+1}L_n^{(\nu+1/2)}(x^2)e^{-x^2/2}\,,
\qquad E_{\nu,n}=
2\nu+4n+3\,, &
\end{eqnarray}
where $E_{\nu,n}$  are the eigenvalues, $n=0,1,\ldots$, and
\begin{eqnarray}
\label{Laguerre}
&L_{n}^{(\alpha)}(x)=\sum_{k=0}^n\frac{\Gamma(n+\alpha+1)}{\Gamma(k+\alpha+1)}
\frac{(-x)^k}{k!(n-k)!}\,&
\end{eqnarray}
are the generalized Laguerre polynomials. 

Then the problem $\mathcal{H}_\nu\psi_\nu=\lambda_\nu\psi_\nu$
fits with (\ref{Sch}), but to generate  new exactly  solvable systems
and identify their ladder operators and 
the associated deformed conformal symmetries 
 we should not be limited 
just by considering  physical states. 
In subsequent sections 
we will 
see how non-physical states obtained via the application of 
certain discrete symmetry group transformations play
an important role in the structure  of the 
quantum conformal algebra generators
of the AFF model
and its  $\mathfrak{osp}(2\vert 2)$ superconformal extensions, as well as
in the construction of new systems and 
their hidden symmetries. 

\subsection{The Klein four-group as a Schr\"odinger equation symmetry}

Parametrizing  the coupling constant in parabolic form 
$g=\nu(\nu+1)$, which is symmetric with respect to $\nu=-\frac{1}{2}$, 
we artificially induce the invariance of the equation 
(\ref{timedependent1}) with respect to the transformation $\rho_1:\nu\rightarrow-\nu-1$.
Equation 
(\ref{timedependent1}) is also invariant 
with respect to the transformation 
$\rho_2:(x,t)\rightarrow(ix,-t)$.
These two transformations generate 
the Klein four-group as a symmetry of equation (\ref{timedependent1}):
$K_4\simeq \Z_2\times \Z_2=(1,\rho_1,\rho_2,\rho_1\rho_2=\rho_2\rho_1)$,
 where each element is its own inverse. 
 At the level of the stationary Schr\"odinger equation,
the action of $\rho_2$ reduces to the transformation
$\rho_2:(x,E_{\nu,n})\rightarrow(ix,-E_{\nu,n})$,
which  means that $\rho_2$
is a completely  broken $\Z_2$ symmetry,
for which  the transformed eigenstates $\rho_2(\psi_{\nu,n})=\psi_{\nu,n}(ix)$ 
 with eigenvalues $-E_{\nu,n}$ are non-physical solutions. 
The transformation
 $\rho_1$ at the same level of the stationary Schr\"odinger equation 
 implies that the energy eigenvalues change as 
 $E_{\nu,n}\rightarrow\rho_1(E_{\nu,n})=E_{-\nu-1,n}=4n-2\nu+1$. 
 The difference between the original energy level and 
 the transformed one is $E_{\nu,n}-E_{-\nu-1,n}=\Delta E\cdot (\nu+1/2)$, 
 where $\Delta E=4$ is the distance between two consecutive levels.
 So, if we take   $\nu=\ell-1/2$ with 
 $\ell=0,1,\ldots$, we obtain 
 $\rho_1(E_{\ell-1/2,n})= E_{\ell-1/2,n-\ell}$,
 and find that physical energy levels with $n\geq\ell$
 are transformed into physical energy levels 
 but lowered by $4\ell$.
 Under  the action of $\rho_1$, the eigenstates  (\ref{physeig})
 are transformed into 
 the functions 
\begin{eqnarray}
\label{psi-nu-1}
 &\rho_1(\psi_{\nu,n})=\sqrt{\frac{n!}{\Gamma(n-\nu+1/2)}}x^{-\nu}L_n^{(-\nu-1/2)}(x^2)e^{-x^2/2}:=\psi_{-\nu-1,n}
 \,.&
\end{eqnarray}
In the case of $\nu\neq \ell-1/2$,
functions (\ref{psi-nu-1}) do not satisfy boundary condition at $x=0$ 
because of the presence of the factor $x^{-\nu}$,
and they are non-physical, formal eigenstates
of $\mathcal{H}_{\nu}$.
The case of $\nu= \ell-1/2$ requires, however,
a separate consideration.
 To analyze this case,  we  observe that
 \begin{eqnarray}
& 
 \rho_1(\psi_{\ell-1/2,n})=\sqrt{\frac{n!}{\Gamma(n-\ell+1)}}x^{-\ell+1/2}L_n^{(-\ell)}(x^2)e^{-x^2/2}\,.
&
 \end{eqnarray}
Due to the poles of Gamma function, this expression vanishes when $n<\ell$, i.e.,
$\rho_1$ annihilates the first $\ell$ eigenstates
of the system.  On the other hand,
the identity 
\be
\frac{(-\eta)^{m}}{m!}L_{n}^{(m-n)}(\eta)=\frac{(-\eta)^{n}}{n!}L_{m}^{(n-m)}(\eta)\,
\ee 
with integer $m$ and $n$, which follows from (\ref{Laguerre}),
allows us to write $\rho_{1}(\psi_{\ell-1/2,n})=(-1)^{\ell}\psi_{\ell-1/2,n-\ell}$ when $n\geq\ell$, 
and 
this is coherent with the change of
the energy eigenvalues under application to them of 
transformation $\rho_1$.
 In conclusion, $\rho_1$ 
corresponds to a  symmetry which 
is just the identity operator when $\ell=0$,  
while for $\ell\geq 1$
this symmetry annihilates the $\ell$ lowest physical eigenstates, 
but restores them by acting on the higher eigenstates~\footnote{This is similar  to a 
picture of a 
Hilbert's  hotel under departure of clients from  first $\ell$  rooms with numbers $n=0,\ldots,\ell-1$ 
with simultaneous  translation of the clients from rooms with numbers $n=\ell,\ell+1,\ldots$, 
into the rooms with numbers $n-\ell$.
Note that the power $(\mathcal{C}^-)^{\ell}$ of lowering generator of conformal symmetry (\ref{Cnu}) 
with $\nu=\ell-\frac{1}{2}$ acts on physical eigenstates in a  way similar to 
$\rho_1$, but violating normalization of the states, see Eq. (\ref{representation1}).}. 
{}From this point of view, in 
the case of half-integer $\nu$, 
transformation $\rho_1$ does not produce anything new.  Nevertheless, we can also 
construct a finite set of non-physical solutions of the same non-physical
nature as in  (\ref{psi-nu-1}) given by the functions
\begin{eqnarray}
\label{halfintegernonphysical}
& \psi_{-\ell-1/2,k}:=
\rho_1\left(\sqrt{\frac{\Gamma(k+l+1)}{k!}}\psi_{\ell-1/2,k}\right)=
x^{-\ell+1/2}L_n^{(-\ell)}(x^2)e^{-x^2/2}, \quad k=0,\ldots,\ell-1,
&
\end{eqnarray}
singular at $x=0$, 
whose corresponding eigenvalues are $E_{-\ell-1/2,n}=4n-2\ell+2$.  
\vskip0.1cm

We note that the combined transformation $\rho_1\rho_2(\psi_{\nu,n})$ 
always produces non-physical solutions for all values of $\nu$
due to the presence of $\rho_2$. 
Wave eigenfunctions transformed by the $K_4$ generators $\rho_2$ 
 and $\rho_1\rho_2$ diverge exponentially at infinity,
 and for the  following consideration it is convenient to introduce 
 a special common notation for them: $\psi_{r(\nu),-n}(x)=\psi_{r(\nu),n}(ix)$,
 where $r(\nu)=\nu$ corresponds to application of $\rho_2$,
 and $r(\nu)=-\nu-1$ corresponds to application of $\rho_1\rho_2$
 to $\psi_{\nu,n}(x)$.
 In the same way, we introduce a common notation for 
 physical eigenstates and non-physical 
eigenfunctions exponentially disappearing 
 at infinity: $\psi_{r(\nu),n}(x)$,
 where $r(\nu)=\nu$ corresponds to eigenstates (\ref{physeig}),
 and $r(\nu)=-\nu-1$ corresponds to 
non-physical eigenfunctions (\ref{psi-nu-1})
or (\ref{halfintegernonphysical})
when $\nu=\ell-1/2$.
In the case of $\nu=\ell-1/2$, $\ell\geq 1$,  we  have
$E_{-\ell-1/2,\ell-n-1}=-E_{-\ell-1/2,n}$ for $n<\ell$, 
and one finds that (\ref{halfintegernonphysical}) and their partners 
in the sense of Eq. (\ref{tildepsi}) are related with 
non-physical eigenstates  produced 
by $\rho_2$ and their partners,   
\be
\label{tilderelation}
\psi_{-\ell-1/2,\ell-1-n}\propto  \widetilde{\psi}_{-\ell-1/2,-n}\,,\qquad 
\widetilde{\psi}_{-\ell-1/2,n}\propto \psi_{-\ell-1/2,-\ell+1-n}\,.
\ee

\subsection{Conformal symmetry and ladder operators}
In this subsection,
we explore the quantum 
conformal 
symmetry of
 the model
 from the perspective of the 
discrete  Klein four-group symmetry.

Hamiltonian (\ref{Lnu}) is the compact generator of the
dynamical conformal  symmetry 
of the AFF model, which together  with 
the second order differential operators 
\begin{eqnarray}
\label{Cnu}
&\mathcal{C}_\nu^-=-\left(\frac{d}{dx}+x\right)^2+\frac{\nu(\nu+1)}{x^2}\,,
\qquad \mathcal{C}_\nu^+=(\mathcal{C}_\nu^-)^\dagger&
\end{eqnarray} 
satisfies the commutation relations of  the 
$\mathfrak{sl}(2,\R)$  algebra,
\be
\label{sl2R}
[\mathcal{H}_{\nu},\mathcal{C}_{\nu}^\pm]=\pm 4 \mathcal{C}_{\nu}^\pm,\qquad 
[\mathcal{C}_{\nu}^-,\mathcal{C}_{\nu}^+]=8\mathcal{H}_{\nu}\,,
\ee
whose Casimir invariant is given by 
$\mathcal{C}=(\mathcal{H}_\nu)^2-\frac{1}{2}(\mathcal{C}_\nu^+\mathcal{C}_\nu^-
+\mathcal{C}_\nu^- \mathcal{C}_\nu^+)=4\nu(\nu+1)-3$.
Taking linear combinations $\mathcal{J}^0:=\frac{1}{4}\mathcal{H}_\nu$,
$\mathcal{J}^1:=\frac{1}{8}(\mathcal{C}_\nu^+ + \mathcal{C}_\nu^-)$,
$\mathcal{J}^2:=\frac{i}{8}(\mathcal{C}_\nu^- - \mathcal{C}_\nu^+)$,
one finds that they satisfy the quantum analog of 
the classical Lorentz algebra 
(\ref{21J}).  The  rescaled Casimir operator $-\frac{1}{16}\mathcal{C}$ represented 
in their terms reduces to $\mathcal{J}^\mu\mathcal{J}_\mu=-\alpha(\alpha-1)$
with $\alpha=\frac{1}{2}\nu+\frac{3}{4}$,
and as it was indicated at the end of the previous section,
 eigenvalues of $\mathcal{J}^0$ are $j^0=n+\alpha$, $n=0,1,\ldots$.
We note here that the linear combinations 
$\mathcal{J}^0-\mathcal{J}^1$ and 
$\mathcal{J}^0+\mathcal{J}^1$ are the 
operators 
$\frac{1}{2}x^2$ and $\frac{1}{2}(-\frac{d^2}{dx^2}+\frac{\nu(\nu+1)}{x^2})$,
which are   the quantum analogs 
of the integrals   $K$ and $H_g$ defined
in (\ref{conformalgenerators})   for the model 
(\ref{conformalaction})  with $q$ changed for $x$ and $t=0$.

Coefficient $4$ in the first commutator in (\ref{sl2R})
is the distance  between the  consecutive 
energy levels of the AFF system, and by means of 
the unitary transformation 
\begin{equation}
\mathcal{C}_{\nu}^{\pm}\rightarrow
 \mathcal{C}_{\nu}^{\pm}(t)=e^{-it\mathcal{H}_\nu}\mathcal{C}_{\nu}^{\pm}e^{it\mathcal{H}_\nu}=e^{\mp i4t}\mathcal{C}_{\nu}^{\pm}\,,
\end{equation}
we obtain two dynamical integrals of motion, now  in the sense of 
the Heisenberg equation 
$\frac{d}{dt}{A}=\frac{\partial A}{\partial t} -i[A,\mathcal{H}_\nu]=0$. 
Their linear Hermitian combinations 
\begin{equation}
\mathcal{D}(t)=i\frac{\mathcal{C}_{\nu}^-(t)-\mathcal{C}_\nu^{+}(t)}{8}\,,\qquad
\mathcal{K}(t)=\frac{2\mathcal{H}_\nu-\mathcal{C}_{\nu}^-(t)-\mathcal{C}_\nu^{+}(t)}{16}\,,
\end{equation}
are the quantum analogs of the generators of the Newton-Hooke symmetry 
 (\ref{NHgenD}) with $\omega=2$.  

According to  (\ref{sl2R}), operators 
$\mathcal{C}_\nu^\pm$ are the ladder operators 
of the quantum AFF system described  by the
Hamiltonian $\mathcal{H}_\nu$. 
The generator $\rho_1$ of the discrete $K_4$ group 
acts identically on 
generators of the conformal symmetry, 
 $\rho_1(\mathcal{H}_\nu)=\mathcal{H}_\nu$,
 $\rho_1(\mathcal{C}_\nu^\pm)=\mathcal{C}_\nu^\pm$,
 while   $\rho_2(\mathcal{H}_\nu)=-\mathcal{H}_\nu$,
  $\rho_2(\mathcal{C}_\nu^\pm)=-\mathcal{C}_\nu^\mp$.
  In correspondence with this,  $\rho_2$ is the automorphism of  the $\mathfrak{sl}(2,\R)$  algebra
  which transforms the unitary irreducible representation $\mathcal{D}^+_\alpha$ 
  of the system (\ref{Lnu})  restricted from below,  
  $j^0=n+\alpha$,   $\alpha=\frac{1}{2}\nu+\frac{3}{4}\geq \frac{1}{2}$,
  into the unitary irreducible representation $\mathcal{D}^-_\alpha$
  of $\mathfrak{sl}(2,\R)$ restricted from above, $j^0=-(n+\alpha)$, $n=0,1,\ldots$ \cite{MPsl}.
  
  The  ladder operators  act on physical eigenstates  and non-physical
states generated from them by
 transformations of the $K_4$ group 
as follows:
\begin{eqnarray}
\label{representation1}
\mathcal{C}_{\nu}^\pm\psi_{r(\nu),n}=\sqrt{(E_{r(\nu),n}\pm2\nu\pm3)(E_{r(\nu),n}\pm2\nu\mp1)}\psi_{r(\nu),n\pm1}\,,\\
\label{representation2}
\mathcal{C}_{\nu}^\pm\psi_{r(\nu),-n}=-\sqrt{(E_{r(\nu),n}\pm2\nu\pm3)(E_{r(\nu),n}\pm2\nu\mp1)}\psi_{r(\nu),-(n\mp1)}\,.
\end{eqnarray}
The coefficients in  (\ref{representation1}) and (\ref{representation2}) 
vanish when the ladder operators act 
on the states from  their corresponding kernels,
 which in the case of $\nu>-1/2$ are given by
\begin{eqnarray}
\label{kerCnu}
\ker\,\mathcal{C}_{\nu}^-=\text{span}\,\{
\psi_{\nu,0},\psi_{-\nu-1,0}
\}\,,\qquad
\ker\,\mathcal{C}_{\nu}^+=\text{span}\,\{
\psi_{\nu,-0},\psi_{-\nu-1,-0}
\}\,.
\end{eqnarray}
In the case of $\nu=-1/2$, the kernels of the ladder operators 
 $\mathcal{C}_{-1/2}^\pm$ 
 are similar to (\ref{kerCnu})  but 
 with  the states  $\psi_{-\nu-1,0}$ and $\psi_{-\nu-1,-0}$ there 
changed, respectively, for  the Jordan states 
\begin{eqnarray}
\label{Jordan0}
&\Omega_{-1/2,0}=\left(a-\frac{1}{2}\ln x\right)\psi_{-1/2,0}\,,\qquad
\Omega_{-1/2,-0}=\left(b-\frac{1}{2}\ln x\right)\psi_{-1/2,-0}\,,&
\end{eqnarray} 
where  $a$ and $b$ are constants. 

In the context of the Darboux transformations, 
the first equation in (\ref{sl2R}) can be written in the equivalent 
form $\mathcal{C}_\nu^\mp\mathcal{H}_\nu=
(\mathcal{H}_\nu\pm 4)\mathcal{C}_\nu^\mp$,
which means that
$\mathcal{C}_\nu^\pm$ intertwine the system $\mathcal{H}_\nu$ 
with itself  but shifted for additive constants $\mp 4$.
Then equation (\ref{kerCnu}) indicates 
that the second order differential operators $-\mathcal{C}_\nu^\pm$ 
are generated by the choice of  the seed states $(\psi_{\nu,\mp0},\psi_{-\nu-1,\pm0})$, and
by means of Eq. (\ref{Darstates})  we can  write the equalities
\begin{eqnarray}
\label{WrC-}
\mathcal{C}_\nu^\mp\phi_{r(\nu),z}=-\frac{W(\psi_{\nu,\pm0},\psi_{-\nu-1,\pm0},
\phi_{r(\nu),z})}{W(\psi_{\nu,\pm0},\psi_{-\nu-1,\pm0})}
\,,
\end{eqnarray}
where $\phi_{r(\nu),z}$ with $z=\pm n$, $n\in \N$, corresponds to an eigenstate or a
Jordan state of $\mathcal{H}_\nu$. 
The Wronskian form of these equalities is useful 
to find  the action of the ladder operators
on the states $\widetilde{\psi}_{r(\nu),\pm 0}$ and $\breve{\Omega}_{-1/2,0}$.  
Using  Eqs. (\ref{ide2}) and (\ref{tech1}), and equalities 
\begin{equation}
\label{tools1}
W(\psi_{\nu,\pm0},\psi_{-\nu-1,\pm0})=-(2\nu+1) e^{\mp x^2}\,,\qquad
W(\psi_{-1/2,\pm0},\Omega_{-1/2,\pm0})=e^{\mp x^2}\,,
\end{equation}
one can find that 
\begin{eqnarray}
\label{tools2}
\mathcal{C}_\nu^-\widetilde{\psi}_{r(\nu),0}\propto\psi_{r(-\nu-1),-0}\,,\qquad
\mathcal{C}_\nu^+\widetilde{\psi}_{r(\nu),-0}\propto\psi_{r(-\nu-1),0}\,,\\
\label{tools3}
\mathcal{C}_{-1/2}^\mp\widetilde{\psi}_{-1/2,\pm0}\propto\Omega_{-1/2,\mp0}\,,\qquad
\mathcal{C}_{-1/2}^\mp\breve{\Omega}_{-1/2,\pm0}\propto\psi_{-1/2,\mp0}\,.
\end{eqnarray}

\begin{figure}[H]
\begin{center}
\includegraphics[scale=0.48]{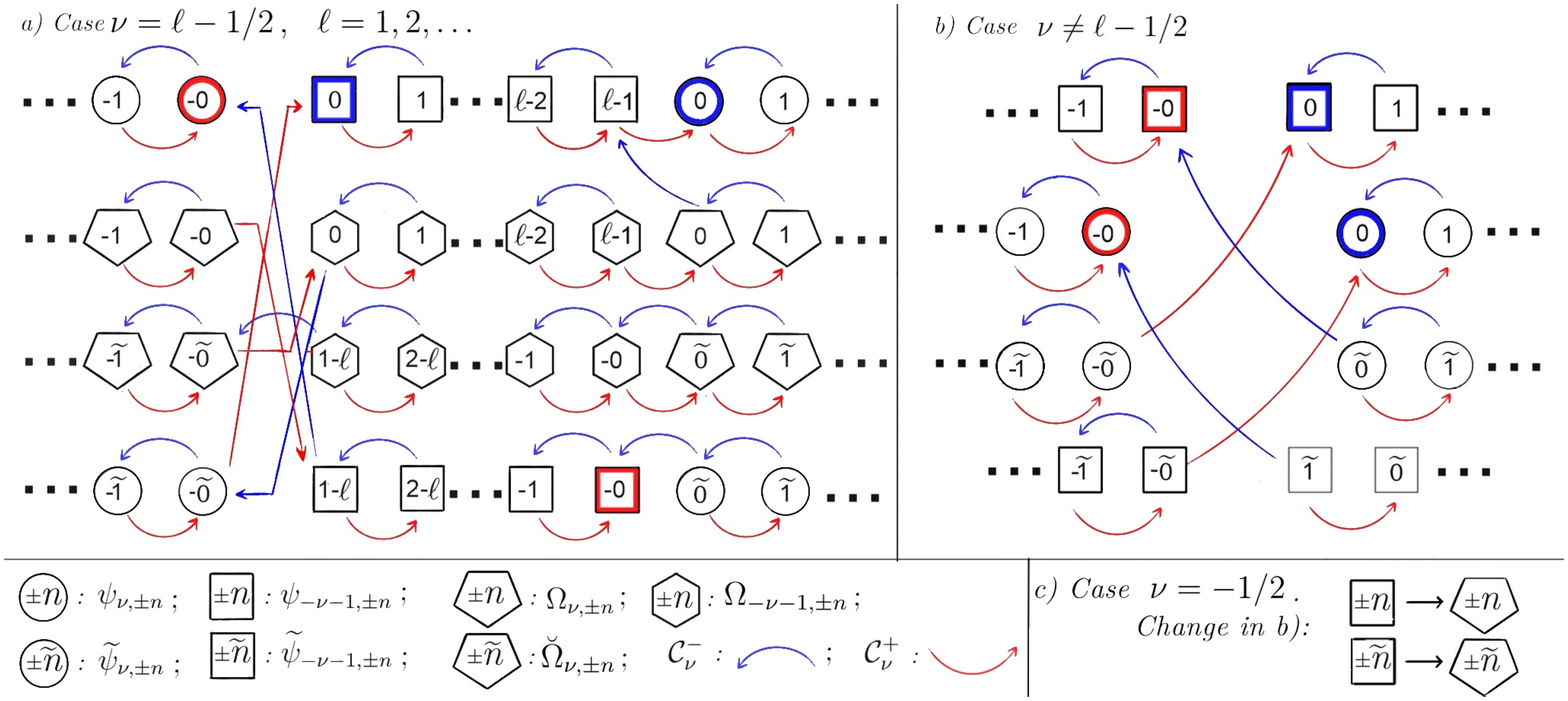} 
\caption{The action of the ladder operators in dependence on the value of $\nu$.
Diagram   \emph{a)} illustrates the case of  half-integer 
$\nu=\ell-1/2$ with $\ell=1,\ldots,$ where it  is shown 
how  Jordan states can be related to eigenstates by the action 
of $\mathcal{C}_{\nu}^\pm$.
Diagram  \textit{b)} corresponds to 
non-half-integer values of  $\nu$.  In \textit{c)}, it is indicated how 
the case with $\nu=-1/2$ 
can be obtained from 
  \textit{b)} by 
changing the corresponding states. 
The shapes with borders highlighted in blue (red) represent  the states
annihilated 
by $\mathcal{C}_{\nu}^-$ ($\mathcal{C}_{\nu}^+$).}
\label{figure0}
\end{center} 
\end{figure}

Returning to the issue of Jordan states, we realize that they  have appeared 
in the systems with half-integer $\nu$, but let us 
consider first the general case. For this, we use (\ref{Dar-jor}) and 
the first relation in (\ref{sl2R}) to prove the relations   
\be
\label{Jordann}
\Omega_{r(\nu),\pm n}\propto(\mathcal{C}_{\nu}^{\pm})^n\Omega_{r(\nu),\pm 0}\,,\qquad
\breve{\Omega}_{r(\nu),\pm n}\propto(\mathcal{C}_{\nu}^{\pm})^n\breve{\Omega}_{r(\nu),\pm 0}\,.
\ee
Thus, the ladder operators act in a similar  way as they act
 on  eigenstates of $\mathcal{H}_\nu$, but with 
a difference when $n=0$.  
When $\nu\not=-1/2$, we obtain the relations
$
\mathcal{C}_{\nu}^{\pm}\Omega_{r(\nu),\mp 0}\propto \widetilde{\psi}_{r(-\nu-1),\pm 0}
$ and $\mathcal{C}_{\nu}^{\pm}\breve{\Omega}_{r(\nu),\mp 0}\propto \Omega_{r(-\nu-1),\pm 0}$.
 Due to  (\ref{tilderelation}) one can make the identification 
 $\breve{\Omega}_{-\ell-1/2,\pm0}=\Omega_{-\ell-1/2,\mp(\ell-1)}$, so in the half-integer case $\nu=\ell-1/2$ 
with $\ell\geq 1$ we obtain
\begin{equation}
\label{ConJordan}
\mathcal{C}^{\pm}_{\ell-1/2}\Omega_{\ell-1/2,\mp0}\propto\psi_{-\ell-1/2,\mp (\ell-1)}\,,\qquad
\mathcal{C}^{\pm}_{\ell-1/2}\Omega_{-\ell-1/2,\mp0}\propto\psi_{\ell-1/2,\mp (\ell-1)}\,.
\end{equation}
 Acting   on  these
 relations by  $(\mathcal{C}_{\ell-1/2}^\pm)^{\ell}$, we obtain zero, and conclude that 
\begin{eqnarray}
\label{spanChalf}
\begin{array}{ll}
\ker (\mathcal{C}_{\ell-1/2}^\pm)^{\ell+k}&=\text{span}\{\psi_{\ell-1/2,\mp0},\ldots,\psi_{\ell-1/2,\mp(\ell+k-1)},\psi_{-(\ell-1/2)-1,\mp0},\ldots,\\
&\qquad\qquad\psi_{-(\ell-1/2)-1,\mp(\ell-1)},\Omega_{\ell-1/2,\mp0},\ldots,\Omega_{\ell-1/2,\mp(k-1)}\}\,
\end{array}
\end{eqnarray} 
for $k=1,2,\ldots$.
The whole picture is summarized in 
Figure   \ref{figure0}.

\subsection{Superconformal symmetry and the Klein four-group}
\label{Section3.2}

Here, we inspect the action of the transformations of the Klein four-group
on a supersymmetric extension of the AFF model.

One can take two different ways
to produce two different supersymmetric extensions of the AFF model by taking 
two different Darboux transformations of the first order
based on the choice of the seed states $\psi_{\nu,0}$ and $\psi_{\nu,-0}$.
By means of Eq. (\ref{generic-inter}) with $n=1$
we obtain two corresponding 
pairs  of Hermitian conjugate intertwining operators,   
\begin{eqnarray}
&
A_{\nu}^-=\frac{d}{dx}+x-\frac{\nu+1}{x}\,,\qquad
A_{\nu}^+=(A_{\nu}^-)^\dagger\,,\qquad
B_{\nu}^-=\frac{d}{dx}-x-\frac{\nu+1}{x}\,,\qquad
B_{\nu}^+=(B_{\nu}^-)^\dagger\,.\qquad&
\label{ABnudef}
\end{eqnarray} 
These operators link the systems $\mathcal{H}_\nu$ and $\mathcal{H}_{\nu+1}$
by the intertwining relations of the form (\ref{inter-gen}),
\begin{eqnarray}
\label{interAnu}
&A_{\nu}^-\mathcal{H}_\nu=(\mathcal{H}_{\nu+1}+2)A_{\nu}^-\,,\qquad
A_{\nu}^+\mathcal{H}_{\nu+1}=(\mathcal{H}_{\nu}-2)A_{\nu}^+\,,&\\&
\label{interBnu}
B_{\nu}^-\mathcal{H}_\nu=(\mathcal{H}_{\nu+1}-2)B_{\nu}^-\,,\qquad
B_{\nu}^+\mathcal{H}_{\nu+1}=(\mathcal{H}_{\nu}+2)B_{\nu}^+\,.&
\end{eqnarray}
Let us  note here that 
if we choose, instead,  non-physical eigenstates $\psi_{-\nu-1,0}$ and
$\psi_{-\nu-1,-0}$ as the seed states,  we 
generate the operators $-A_{\nu-1}^+$ 
and $-B_{\nu-1}^+$, respectively. 
These operators relate the system $\mathcal{H}_{\nu}$ with
$\mathcal{H}_{\nu-1}$ by intertwining relations of the form 
(\ref{interAnu}) and  (\ref{interBnu}) with $\nu$ shifted by minus one.

{}From the point of view of the Klein four-group,
the states  $\psi_{\nu,0}$ and $\psi_{\nu,-0}$ are related 
by transformation $\rho_2$, while  the states 
$\psi_{\nu,\pm0}$ and $\psi_{-\nu-1,\pm0}$ are mutually transformed  by $\rho_1$. 
As a consequence,
the application of transformations $\rho_1$ and $\rho_2$
to the intertwining operators gives
\begin{eqnarray}
\label{rho1}
\rho_1(A^\mp_{\nu})=-B^\pm_{\nu-1}\,,\qquad
\rho_1(B^\mp_{\nu})=-A^\pm_{\nu-1}\,,\\
\label{rho2}
\rho_2(A^\pm_\nu)=-iB^\pm_\nu\,,\qquad
\rho_2(B^\pm_\nu)=-iA^\pm_\nu \,.
\end{eqnarray}
These relations are valid for $\nu>-1/2$.  At 
 $\nu=-1/2$ the transformation $\rho_1$ reduces to the identity. 

We can construct now two different $\mathcal{N}=2$ super-Hamiltonians 
\begin{eqnarray}
\label{hamiliso}
\mathcal{H}_\nu^{e}=
\left(
\begin{array}{cc}
A_\nu^-A_\nu^+=\mathcal{H}_{\nu+1}-2\nu-1&    0 \\
0 & A_\nu^+A_\nu^-= \mathcal{H}_{\nu}-2\nu-3    
\end{array}
\right)\,,\\
\label{hamilisob}
\mathcal{H}_\nu^{b}=
\left(
\begin{array}{cc}
B_\nu^-B_\nu^+=\mathcal{H}_{\nu+1}+2\nu+1&    0 \\
0 & B_\nu^+B_\nu^-= \mathcal{H}_{\nu}+2\nu+3    
\end{array}
\right)\,,
\end{eqnarray}
where the indexes $e$ and $b$ refer to the 
exact and spontaneously broken $\mathcal{N}=2$ Poincar\'e supersymmetries
described by the corresponding super-Hamiltonians.
The operator      
(\ref{hamiliso}) has an equidistant spectrum 
 given by eigenvalues $\mathcal{E}_n=4n,$  $n=0,1,\ldots$,
where $n=0$ corresponds to the nondegenerate ground state 
$(0,\psi_{\nu,0})^t$ of zero energy, 
while  all the energy levels with $n\geq 1$ are doubly degenerate.
The Hamiltonian (\ref{hamilisob}) has eigenvalues
 $\mathcal{E}_n=4n+4\nu+6$, each  of which is doubly degenerate,
 and   two ground states with energy  $\mathcal{E}_0=4\nu+6>0$ are
$\Psi_{0}^{(+)}=(B_\nu^-\psi_{\nu-1,0},\psi_{\nu-1,0})^t$ 
and  
$\Psi_{0}^{(-)}=\sigma_3\Psi_{0}^{(+)}$.

System (\ref{hamiliso}) is described by the $\mathfrak{osp}(2,2)$ superconformal 
dynamical symmetry generated by the even,
$\mathcal{H}_\nu^{e}$, $\mathcal{R}_\nu=\frac{1}{4}(\mathcal{H}_\nu^{e}-\mathcal{H}_\nu^{b})=
\frac{\sigma_3}{2}-(\nu+1)\mathbb{I}$,
$\mathcal{G}_\nu^\pm$,
and odd, 
$\mathcal{Q}_\nu^a$,
$\mathcal{S}_\nu^a$, $a=1,2$, operators, 
where $\mathbb{I}$ is the unit $2\times 2$ matrix, 
\begin{eqnarray}\label{lader}
&\mathcal{G}_\nu^{\pm}=
\left(
\begin{array}{cc}
\mathcal{C}_{\nu+1}^\pm&   0  \\
 0 &  \mathcal{C}_{\nu} ^\pm    
\end{array}
\right),&\\
\label{supercharge1}
&\mathcal{Q}_{\nu}^{1}=
\left(
\begin{array}{cc}
  0&    A^-_\nu  \\
 A^+_\nu &   0     
\end{array}
\right),
\qquad
 \mathcal{S}_\nu^1=
\left(
\begin{array}{cc}
  0&    B^-_{\nu}  \\
 B^+_{\nu} &   0     
\end{array}
\right),&\\
&\mathcal{Q}_\nu^{2}=
i\sigma_3\mathcal{Q}_\nu^1\,,
\qquad \mathcal{S}_\nu^2=
i\sigma_3\mathcal{S}_\nu^1\,.&
\end{eqnarray}
Here $\mathcal{Q}_\nu^{a}$ are the supercharges of the system, which annihilate
the ground state, i.e, 
the system indeed is in  the  phase of unbroken 
$\mathcal{N}=2$ Poincar\'e
supersymmetry. 
The Lie superalgebraic relations 
\begin{eqnarray}\label{HRQ0}
&[\mathcal{H}_\nu^{e},\mathcal{R}_\nu]=[\mathcal{H}_\nu^{e},\mathcal{Q}_\nu^a]=0\,,&\\
\label{evencommutation}
&[\mathcal{H}_\nu^{e},\mathcal{G}_\nu^{\pm}]=\pm4\mathcal{G}_\nu^{\pm}\,, \qquad 
[\mathcal{G}_\nu^{-},\mathcal{G}_\nu^{+}]=8\mathcal{H}^{e}_\nu-16\mathcal{R}_\nu\,,&\\
\label{evenodd}
&[\mathcal{H}_\nu^{e},\mathcal{S}_\nu^a]=-4i\epsilon^{ab}\mathcal{S}_\nu^b\,,\qquad
[\mathcal{R}_\nu,\mathcal{Q}_\nu^a]=-i\epsilon^{ab}\mathcal{Q}_\nu^b\,,
\qquad
[\mathcal{R}_\nu,\mathcal{S}_\nu^a]=-i\epsilon^{ab}\mathcal{S}^b_\nu\,,&\\
\label{fq1}
&[\mathcal{G}_\nu^-,\mathcal{Q}_\nu^a]=2(\mathcal{S}_\nu^a+i\epsilon^{ab}\mathcal{S}_\nu^b), \qquad 
[\mathcal{G}_\nu^+,\mathcal{Q}_\nu^a]=-2(\mathcal{S}_\nu^a-i\epsilon^{ab}\mathcal{S}_\nu^b)\,,&\\
\label{fq3}
&[\mathcal{G}_\nu^-,\mathcal{S}_\nu^a]=2(\mathcal{Q}_\nu^a-i\epsilon^{ab}\mathcal{Q}_\nu^b)\,, \qquad 
[\mathcal{G}_\nu^+,\mathcal{S}_\nu^a]=-2(\mathcal{Q}_\nu^a+i\epsilon^{ab}\mathcal{Q}_\nu^b)\,,&\\
\label{anti1}
&\{ \mathcal{Q}_\nu^a,\mathcal{Q}_\nu^b\}=2\delta^{ab}\mathcal{H}_\nu^{e}\,, \qquad 
\{ \mathcal{S}_\nu^a,\mathcal{S}_\nu^b\}=2\delta^{ab}(\mathcal{H}_\nu^{e} -4\mathcal{R}_\nu)\,,&\\
\label{anti2}
&\{\mathcal{Q}^a_\nu,\mathcal{S}^b_\nu\}=\delta^{ab}(\mathcal{G}_\nu^{+}+\mathcal{G}_\nu^-)+
i\epsilon^{ab}(\mathcal{G}_\nu^+-\mathcal{G}_\nu^-)\,&\label{QSGG}
\end{eqnarray}
correspond to the  dynamical $\mathfrak{osp}(2,2)$ 
superconformal symmetry of the system 
(\ref{hamiliso}).
To identify the generators of superconformal symmetry of the system
(\ref{hamilisob}),
we note that  
the supeconformal $\mathfrak{osp}(2,2)$ algebra given
by Lie super-algebraic relations (\ref{HRQ0})--(\ref{QSGG})
 has an automorphism 
$f=f^{-1}$
which corresponds to  the transformations 
$\mathcal{H}_{\nu}^{e}\rightarrow \mathcal{H}_{\nu}^{e}-4\mathcal{R}_{\nu}=\mathcal{H}_{\nu}^b$,
$\mathcal{R}_{\nu}\rightarrow -\mathcal{R}_\nu$, 
$\mathcal{G}_{\nu}^\pm\rightarrow \mathcal{G}_{\nu}^{\pm}$,
$\mathcal{Q}_\nu^1\rightarrow -\mathcal{S}_{\nu}^{1}$, 
$\mathcal{Q}_\nu^2\rightarrow \mathcal{S}_{\nu}^{2}$,
$\mathcal{S}_\nu^1\rightarrow -\mathcal{Q}_{\nu}^{1}$
$\mathcal{S}_\nu^2\rightarrow \mathcal{Q}_{\nu}^{2}$.
Therefore, the transformed operators are the generators of the 
$\mathfrak{osp}(2,2)$ corresponding to the system 
$\mathcal{H}_{\nu}^b$, for which 
the Poincar\'e supercharges are  $-\mathcal{S}_\nu^1$ and 
$\mathcal{S}_{\nu}^2$. None of these 
supercharge operators annihilates both
ground states of the system
coherently with spontaneously broken nature  
of the $\mathcal{N}=2$ Poincar\'e
supersymmetry of the system (\ref{hamilisob}).

Let us consider now  the action of the $K_4$ group on the 
super-extended systems.
The action of both generators $\rho_1$ and $\rho_2$ 
on super-extended systems we define in the same way 
as they were defined for the non-extended AFF system 
$\mathcal{H}_\nu$. 
We use Eq.  (\ref{rho1})
to transform  the  generators of $\mathfrak{osp}(2,2)$  by $\rho_1$, and  
obtain
\begin{eqnarray}
\label{gentransformed1}
&\rho_1(\mathcal{H}_{\nu}^{e})=\sigma_1(\mathcal{H}_{\nu-1}^{e}-4\mathcal{R}_{\nu-1})\sigma_1\,,\qquad
\rho_1(\mathcal{G}_{\nu}^\pm)=\sigma_1(\mathcal{G}_{\nu-1}^\pm)\sigma_1\,,&\\
&\rho_1(\mathcal{R}_{\nu})=\sigma_{1}(-\mathcal{R}_{\nu-1})\sigma_{1}\,,&\label{gentransformed1+}\\
&\rho_1(\mathcal{Q}_{\nu}^1)=\sigma_1(-\mathcal{S}_{\nu-1}^1)\sigma_1\,,\qquad
\rho_1(\mathcal{Q}_{\nu}^2)=\sigma_1(\mathcal{S}_{\nu-1}^2)\sigma_1\,,\label{gentransformed1++}& \\
&\rho_1(\mathcal{S}_{\nu}^1)=\sigma_1(-\mathcal{Q}_{\nu-1}^1)\sigma_1\,,\qquad
\rho_1(\mathcal{S}_{\nu}^2)=\sigma_1(\mathcal{Q}_{\nu-1}^2)\sigma_1\,.\label{gentransformed2}&
\end{eqnarray}
Therefore, the action of $\rho_1$ on generators 
of the superconformal algebra
of the system described by the super-Hamiltonian $\mathcal{H}_{\nu}^{e}$
produces the generators of supeconformal symmetry 
$\mathfrak{osp}(2,2)$ of the system described by the super-Hamiltonian
 $\mathcal{H}_{\nu-1}^{b}$ unitarily rotated by
 $\sigma_1$.
 In other words, the action of $\rho_1$ on the system (\ref{hamiliso})
 and generators of its superconformal algebra
is equivalent to do the shift $\nu\rightarrow\nu-1$, apply 
the automorphism $f$ defined above, 
and then apply a unitary transformation given by the operator 
$\sigma_1$. 
The transformed generators 
(\ref{gentransformed1})-(\ref{gentransformed2}) 
still satisfy the same superconformal algebra, 
i.e. $\rho_1$ is an automorphism of the $\mathfrak{osp}(2,2)$,
but they describe another
super-extended system having  different 
properties in the sense that in the transformed system,
unlike the  initial system (\ref{hamiliso}),
the $\mathcal{N}=2$ Poincar\'e  supersymmetry 
is  spontaneously broken in the case of $\nu>-1/2$. The only  
exception from this rule corresponds to 
the case $\nu=-1/2$, where the the transformed Hamiltonian
reduces to 
$\sigma_1\mathcal{H}_{-1/2}^{e}\sigma_1$, 
and represents a unitary transformed 
super-Hamiltonian with the unbroken 
 $\mathcal{N}=2$ Poincar\'e  supersymmetry.

On the other hand, one can verify that when $\rho_1$
 acts on the Hamiltonian
$\mathcal{H}_\nu^{b}$, it produces $\sigma_1(\mathcal{H}_{\nu-1}^{e})\sigma_1$, 
and this time  the  $\mathcal{N}=2$ Poincar\'e  supersymmetry
 of the system is changed from the
spontaneously broken phase (in the case of $\nu>-1/2)$
to the phase of unbroken supersymmetry,
 with the only exception of the system $\mathcal{H}_{-1/2}^b$
 with unbroken supersymmetry,
which unitary transforms into  $\sigma_1\mathcal{H}_{-1/2}^{b}\sigma_1$.  
This picture of  transformation of $\rho_1$ on super-extended 
systems can be compared with 
the case of the non-extended AFF system, where $\rho_1$
acts identically on its Hamiltonian  and generators of the conformal symmetry,
though, as we saw, it acts  nontrivially  on eigenstates of the system.

By knowing the action of  
$\rho_2$ on
the intertwining operators (\ref{ABnudef}), explicilty given in (\ref{rho2}), 
we can apply this transformation to  the generators of superconformal 
algebra.
 As a result, we find that 
it generates the automorphism of  
$\mathfrak{osp}(2,2)$ given by relations
\begin{eqnarray}
\label{gentransformedrho2}
&\qquad \rho_2(\mathcal{H}_{\nu}^{e})= -\mathcal{H}_{\nu}^{b}\,,\quad
\rho_2(\mathcal{G}_{\nu}^\pm)= -\mathcal{G}_{\nu}^{\mp}\,,\quad
\rho_2(\mathcal{R}_{\nu})= \mathcal{R}_\nu\,,&\\
&\rho_2(\mathcal{Q}_\nu^1)= -i\mathcal{S}_{\nu}^{1}\,, \qquad
\rho_2(\mathcal{Q}_\nu^2)= -i\mathcal{S}_{\nu}^{2}\,,&\\
&\rho_2(\mathcal{S}_\nu^1)= -i\mathcal{Q}_{\nu}^{1}\,,\qquad
\rho_2(\mathcal{S}_\nu^2)= -i\mathcal{Q}_{\nu}^{2}\,.&
\end{eqnarray}
Transformed Hamiltonian operator is similar here to the Hamiltonian  
produced  by the automorphism $f$  but  multiplied by $-1$.
This correlates with the anti-Hermitian nature of 
the transformed fermion generators of superalgebra.
Accordingly,  the spectrum of the transformed matrix Hamiltonian
is  negative, not bounded from below, and each its level is doubly degenerate 
for  $\nu\geq-1/2$.

In correspondence with the described picture, 
the application of the combined transformation $ \rho_2\rho_1$
is just another automorphism of the superconformal algebra
(\ref{HRQ0})-(\ref{anti2}), which produces 
anti-Hermitian  odd generators,  and 
$\rho_2\rho_1(\mathcal{H}_{\nu}^{e})=\sigma_1(-\mathcal{H}_{\nu-1}^{e})\sigma_1$.
The discrete spectrum of  the transformed Hamiltonian
is not restricted from below and is  given
by the numbers $\mathcal{E}_n=-4n$, $n=0,1,\ldots$,
where each negative energy level is doubly degenerate, 
while non-degenerate zero energy level corresponds to the 
state  $(\psi_{\nu,0},0)^t$.

\section{Dual Darboux schemes}
\label{Mirror}
%%%%%%%%%%%%%%%%%

If we have two ways to generate the same, modulo an additive shift,
 particular system from 
the AFF model with a certain value of the parameter $\nu$ by employing 
two distinct  Darboux transformations based on distinct sets of the seed states
with  different behavior at infinity,
we say that we have two dual Darboux schemes, or a Darboux duality. 
This property was  discussed and exploited earlier  in the case of 
rational deformations of harmonic oscillator for the construction
of the complete sets of the spectrum generating ladder operators
as well as for the description of emergent nonlinear extensions
of superconformal symmetry appearing in such systems
 \cite{CarPly,CarPly2,InzPly,CIP,LM2}.  
 We also used dual Darboux schemes in the previous 
 section in the discussion of supersymmetric extensions
 of the AFF model.

To give a simple example of the dual schemes,
one can choose the set of eigenstates $(\psi_{\nu,0},\ldots,\psi_{\nu,m-1})$, 
whose Wronskian is, up to a multiplicative constant, 
$x^{m(2\nu+m+1)/2}e^{-mx^2/2}$. This implies that after the
Darboux transformation  (\ref{Dar}) the Hamiltonian operator 
takes the form $\mathcal{H}_{\nu+m}+2m$. On the other hand, if we take the scheme   
$(\psi_{\nu,-0},\ldots,\psi_{\nu,-(m-1)})$ based on non-physical 
eigenstates generated by transformation $\rho_2$, we produce the same 
system but shifted by the relative constant $-4m$, so this both schemes are dual.
Intertwining operators of order $m$
of both schemes allow to construct fermionic integrals of motion 
of the corresponding supersymmetric extension of the  system, 
and  generate the corresponding $\mathfrak{osp}(2,2)$ superconformal symmetry
in the case of $m=1$, as it was done
in the previous section,  or 
to generate non-linearly deformed and extended version  of $\mathfrak{osp}(2,2)$ 
when $m>1$.

The purpose of this section is to derive certain Wronskian identities 
 and employ them for construction of  the  dual Darboux schemes.
 The latter allow us to generate rational deformations of a general form 
with arbitrary number of gaps implemented into  equidistant spectrum of the AFF 
systems in a well controlled manner, to identify the complete sets of the spectrum generating ladder 
operators for them and their nonlinearly deformed conformal symmetries.
We also observe the peculiar changes that happen with spectra of such
systems under variation of the parameter $\nu$ when 
it goes through half-integer values. 
 
In the following the equalities between wave functions and Wronskians 
are considered up to multiplication by nonzero real constants.

\subsection{The dual schemes algorithm}

\label{DualSec}
Here we  consider a systematic procedure
 to obtain the dual scheme from a given Darboux scheme
by means of certain Wronskian identities. 

As we have shown 
in the previous section, the case in which  $\nu$ takes half-integer values 
 is special  and more complicated in comparison with  the non-half-integer case. 
This happens due to appearance of Jordan states in the game 
via the  properties   of  non-compact generators of the 
conformal symmetry which simultaneously are the ladder operators  
for corresponding AFF systems, see Eq.  (\ref{ConJordan}).
By this reason we start first with the case of non-half-integer values of $\nu$,  and 
choose a generic set of physical and non-physical eigenstates
of $\mathcal{H}_{\nu}$ as  the seed states,
\begin{eqnarray}
\label{unioncollection}
\{\alpha\}=(\psi_{\nu,k_1},\ldots,\psi_{\nu,k_{N_1}},\psi_{-\nu-1,l_1},\ldots,\psi_{-\nu-1,l_{N_2}})
\,,\qquad
k_{i},l_{j}= \pm0,\pm1,\ldots\,,
\end{eqnarray}
where $i=1,\ldots,N_1$ and $j=1,\ldots,N_2$, and, 
for simplicity,
we suppose that  
$|k_1|<\ldots<|k_{N_1}|$ and  $|l_1|<\ldots<|l_{N_2}|$.

Consider now a scheme of the form (\ref{unioncollection}) with 
non-repeated states, and suppose that 
both $k_i$ and $l_j$ carry the same sign for all $i$ and $j$.
Also let us define the index number $n_N=\text{max}\,(|k_1|,\ldots,|k_{N_1}|,|l_1|,\ldots,|l_{N_2}|)$,
which can correspond to a state with index $\nu$ or $-\nu-1$. 
So if $k_i$ and $l_j$ carry the plus sign, the equality  
\begin{eqnarray}
\label{eqschemes3}
&W(\{\alpha\})=e^{-(n_{N}+1)x^2}W(\{\Delta_-\})\,,&\\\nonumber
&\{\Delta_-\}:=(\psi_{-\nu-1,-0},\psi_{\nu,-0},\ldots,\check{\psi}_{-\nu-1,-r_i},
\check{\psi}_{\nu,-s_i},\ldots,\psi_{-\nu-1,-n_N},{\psi}_{\nu,-n_N}
)\,,&
\end{eqnarray}   
is satisfied, where the marked  states
 $\check{\psi}_{-\nu-1,-r_i}$ and $\check{\psi}_{\nu,-s_i}$, with 
 $r_i=n_{N}-k_i$ and $s_j=n_{N}-l_j$,
are omitted 
from the set $\{\Delta_-\}$.
On the contrary, if $k_i$ and $l_j$ carry the minus sign, we have the equality
\begin{eqnarray}
\label{eqschemes4}
&W(\{\alpha\})=e^{(n_{N}+1)x^2}W(\{\Delta_+\})\,,&\\\nonumber
&\{\Delta_+\}:=(\psi_{-\nu-1,0},\psi_{\nu,0},\ldots,\check{\psi}_{-\nu-1,r_i},
\check{\psi}_{\nu,s_j},\ldots,{\psi}_{-\nu-1,n_N},{\psi}_{\nu,n_N}
)\,,&
\end{eqnarray}   
where now $r_i=n_N-|k_i|$ and $s_j=n_N-|l_j|$.
These relations are also valid if one of the
numbers ${N_1}$ or ${N_2}$
is equal to zero, which means that in the corresponding scheme there are only 
states of the same kind
with respect to the first index,  $-\nu-1$ or $\nu$, respectively. 

When considering  $\nu=\ell-1/2$ with $\ell=0,1,2$,
 we will have  relations analogous to (\ref{eqschemes3}) and 
(\ref{eqschemes4}),  but  changing each state of the form 
$\psi_{-\nu-1,\pm(\ell+k)}$ by $\Omega_{\ell-1/2,\pm k}$,
as a result 
we deal with the confluent Darboux transformation.

To prove the displayed  Wronskian relations,
we have to keep 
in mind that the operators 
$\mathcal{C}_{\nu}^\pm$ 
 are in fact the intertwining operators 
of the dual schemes 
$(\psi_{\nu,0},\psi_{-\nu-1,0})$ and
$(\psi_{\nu,-0},\psi_{-\nu-1,-0})$ in the case of $\nu>1/2$,
while for $\nu=-1/2$ we take the sets   $(\psi_{1/2,0},\Omega_{-1/2,0})$ and 
$(\psi_{1/2,-0},\Omega_{-1/2,-0})$.
It is also necessary to take into account the action of these operators 
on eigenstates  and Jordan states discussed  in the previous  section.

By means of  equations (\ref{id1}), (\ref{ide2}), (\ref{tools1}) and (\ref{tools3}), 
we will develop a step by step 
processes to show the validity of  (\ref{eqschemes3}) and (\ref{eqschemes4})
in the case when $\nu$ is not a half-integer number, and then we 
will explain how 
these relations  can be extended to the half-integer case.      
As a starting point,  consider the Wronskian of the set 
$\{\alpha\}$  
 defined in (\ref{unioncollection}).  
If the states 
 $\psi_{\nu,\pm 0}$ and $\psi_{-\nu-1,\pm 0}$ do not belong to (\ref{unioncollection}),
 we can replace the Wronskian $W(\{\alpha\})$ by 
\begin{eqnarray}
\label{updown1}
&W(\psi_{\nu,\pm0},\psi_{-\nu-1,\pm0},\widetilde{\psi}_{\nu,\pm0},\widetilde{\psi}_{-\nu-1,\pm0},\{\alpha\})=
e^{\mp x^2}W(\psi_{\nu,\mp0},\psi_{-\nu-1,\mp0},\{\mathcal{C}_\nu^\mp \alpha\})\,,&
\end{eqnarray}
where we used relations (\ref{id1}), (\ref{ide2}), (\ref{tools1}) and
(\ref{tools2}), and  $\{\mathcal{C}_\nu^\mp\alpha\}$ means 
that the ladder operators are applied to  all the states in the set. 
On the other hand, if $\psi_{r(\nu),\pm 0}$ belong to (\ref{unioncollection}),
we can replace the Wronskian of the initial set of the seed states by 
\begin{eqnarray}
\label{updown2}
W(\psi_{r(-\nu-1),\pm0},\widetilde{\psi}_{r(-\nu-1),\pm0},\{\alpha\})=
e^{\mp x^2}W(\psi_{r(-\nu-1),\mp0},\{\mathcal{C}_\nu^\mp\beta_1\})\,,
\end{eqnarray}
where $\{\beta_1\}$ is the scheme $\{\alpha\}$ with the omitted state $\psi_{r(\nu),\pm0}$. 
Finally, if  $\psi_{\nu,\pm 0}$ and $\psi_{-\nu-1,\pm 0}$ belong to (\ref{unioncollection}),
we have
\begin{eqnarray}
\label{updown3}
&W(\{\alpha\})=e^{\mp x^2}W(\{\mathcal{C}_\nu^\mp\beta_2\})\,,&
\end{eqnarray}
where $\{\beta_2\}$ is the scheme $\{\alpha\}$ with the omitted states 
$\psi_{\nu,\pm0}$ and $\psi_{-\nu-1,\pm0}$.  
Note that in all these three relations  we have lowered or raised the index 
of the states in $\{\alpha\}$, 
and also in the case of Eqs.  (\ref{updown1}) and (\ref{updown2})
we have included additional states which do not belong to the initial set. 
Also, we  note that an exponential factor has appeared. 
These identities can be applied to the Wronskians  on the right hand side of  equations 
(\ref{updown1})-(\ref{updown3}), 
which will contribute with new exponential factors in new Wronskians, and so on.
For this reason,  if we restrict the initial set $\{\alpha\}$ by the conditions 
described above (that every state in the set has the second index of the same sign), 
and we repeat this procedure $n_N+1$ times
with positive (negative) sign of the indexes 
in (\ref{updown1})-(\ref{updown3}), 
we finally obtain  equation (\ref{eqschemes3}) or (\ref{eqschemes4}).

Now, we consider the case when $\nu$ takes a half-integer value. 
Note first that 
by means of relations (\ref{id1}), (\ref{tech1}), the second relation 
in (\ref{tools1}), and  relations (\ref{tools3}), we can repeat 
the arguments presented above for the case when $\nu=-1/2$,
 but changing each function of the form $\psi_{-\nu-1,n}$ by 
$\Omega_{-1/2,n}$ in relations (\ref{updown1})-(\ref{updown3}).  As a  consequence,
relations  (\ref{eqschemes3}) and (\ref{eqschemes4}) are now valid with the 
same corresponding changes. 
On the other hand,  
with a simple example one
can see that this does not hold for the case 
$\nu=\ell-1/2$ with $\ell\geq 1$. 
For this  we  consider the scheme  $(\psi_{1/2,1},\psi_{1/2,2})$, 
for which the Wronskian can be rewritten as 
\begin{equation}
W(\psi_{1/2,0},\psi_{3/2,0},\widetilde{\psi}_{1/2,0},\widetilde{\psi}_{-3/2,0},\psi_{1/2,1},\psi_{1/2,2})=
e^{-x^2}W(\widetilde{\psi}_{-3/2,0},\psi_{1/2,-0},\psi_{1/2,0},\psi_{1/2,1})\,,
\end{equation}
where we have repeated the same idea that we employed in (\ref{updown1}), 
and also we used (\ref{tilderelation}) to change $\psi_{-3/2,-0}$ 
by $\widetilde{\psi}_{-3/2,0}$. As this last indicated state appears, 
we cannot use equations
(\ref{id1}) and (\ref{ide2}) to include 
 $\psi_{-3/2,0}$ and produce the intertwining operator $\mathcal{C}_{1/2}^-$, so
 the algorithm is stopped.

Nevertheless, we can use  the 
 connection between $\mathcal{H}_{-1/2+\ell}$
and  $\mathcal{H}_{-1/2}$, provided by the Darboux transformation produced 
by the seed states $(\psi_{-1/2,\pm0},\ldots,\psi_{-1/2,\pm(\ell-1)})$ to obtain 
the corresponding dual schemes.
Each eigenstate  or Jordan state of $\mathcal{H}_{\nu+m}$ can be obtained 
by applying the Darboux mapping to corresponding 
eigenstates or Jordan states 
of $\mathcal{H}_{-1/2}$.   
The details of the procedure are
described  
in Appendix \ref{dualgamma}, but it can be
summarized in three simple steps. 
\begin{itemize}
\item If we have a scheme based on  eigenstates or Jordan 
states of $\mathcal{H}_{\nu+m}$
with $\nu=\ell-1/2$,
 then by using the corresponding Darboux transformation and Eq. (\ref{id1}), 
we find an equivalent scheme in the system 
with $\nu=-1/2$.
\item Then we construct the dual scheme by using the algorithm
adapted  for $\nu=-1/2$.

\item Finally, we translate the resulting dual scheme into the scheme  
for
$H_{\nu+m}$ by using the corresponding 
Darboux transformation and equation (\ref{id1}) again. 
\end{itemize}
The main result 
 is that 
we just have to change every function of the form $\psi_{-\nu-1,\pm(\ell+n)}$
 by $\Omega_{-\ell-1/2,\pm n}$ when  $\nu$ is equal to  $\ell-1/2$. 
 In this way  one finds  that 
  $W(\psi_{1/2,1},\psi_{1/2,2})=e^{-3x^2}W(\psi_{1/2,-0},\psi_{1/2,-1},\psi_{1/2,-2},\Omega_{1/2,-1})$. 
  \vskip0.1cm

Now, we focus our discussion on 
relation between both dual schemes. 
In general, if $\{\Delta_-\}$ coincides with the scheme in the argument of
the Wronskian on the left hand side of  
(\ref{eqschemes4}), then $\{\Delta_+\}$ coincides with the scheme on the left hand side of (\ref{eqschemes3}), 
and consequently $W(\{\Delta_+\})=e^{-(n_N+1)x^2}W(\{\Delta_-\})$.
By this reason 
we call $\{\Delta_+\}$ and $\{\Delta_-\}$
a positive and a negative dual scheme, respectively. 
If $\{\Delta_+\}$ has $n_+$
states and $\{\Delta_-\}$ has $n_-$ states, then
 one can note that $n_N=n_{n_+}=n_{n_-}$,
  and $n_++n_-=2(n_N+1)$, which is the total number of 
the states employed in both dual transformations.

\begin{figure}[H]
\begin{center}
\includegraphics[scale=0.31]{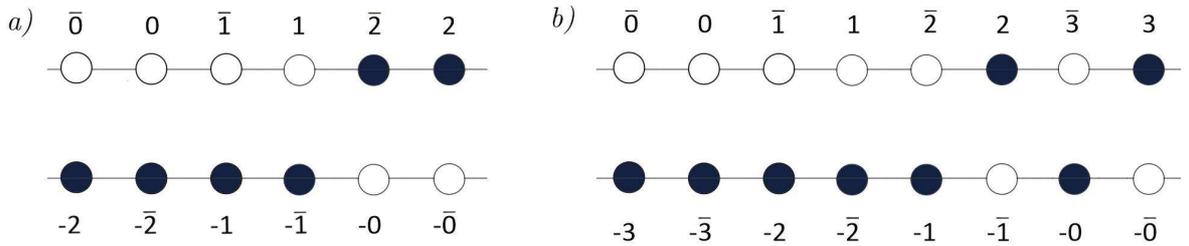} 
\caption{Two ``mirror diagrams" corresponding to dual schemes 
for the conformal mechanics model. 
The numbers $n$ indicate the states $\psi_{\nu,n}$, and symbols 
$\bar{n}$ correspond 
to the  states $\psi_{-\nu-1,n}$.} 
\label{figure1}
\end{center} 
\end{figure}

The general  
rules can be summarized and better understood with the 
examples presented diagrammatically  in Fig. \ref{figure1}. 
This kind of diagrams is read as follows. In the top-line, there appear the ordered 
states vanishing at infinity, which are ordered  from the lowest to the highest second index
in wave functions, and which always end in the number without a bar
(the first index of wave function is $\nu$). In the bottom-line, there 
appear the Wick rotated states (second index of wave functions appears with the minus sign), 
ordered in the same way. The filled black circles denote the states
that appear in the Wronskian arguments in the corresponding  dual scheme. 
The mirror diagrams  shown in Fig. \ref{figure1} 
correspond to  the following Wronskian equalities:  
\begin{eqnarray}
\label{Interpoeg}
W(\psi_{-\nu-1,2},\psi_{\nu,2})=e^{-3x^2}W(\psi_{-\nu-1,-1},\psi_{\nu,-1},\psi_{-\nu-1,-2},\psi_{\nu,-2})\,,\\
\label{Adlerkreineg}
W(\psi_{\nu,2},\psi_{\nu,3})=e^{-4x^2}W(\psi_{\nu,-0},\psi_{\nu,-1},\psi_{-\nu-1,-2},\psi_{\nu,-2},,\psi_{-\nu-1,-3},\psi_{\nu,-3})\,,
\end{eqnarray}
whose  explicit form is given in Appendix \ref{Apendix-Wronskian}.
The transformation which relates the AFF systems 
described by  $\mathcal{H}_\nu$ with $\mathcal{H}_{\nu+m}$ 
can also be understood within  this picture. Furthermore, using 
a diagram similar to those  in Fig. \ref{figure1}, one can 
show that the schemes 
$\{\Delta_+\}=(\psi_{r(\nu),0},\ldots,\psi_{r(\nu),m-1})$ and 
$\{\Delta_-\}=(\psi_{r(\nu),-0},\ldots,\psi_{r(\nu),-(m-1)})$ are dual.

%============

\section{Rationally deformed AFF systems}
\label{Ladders}

A rational deformation of the AFF model can be generated 
by taking a set of the seed states
\begin{equation}\label{alpKA}
\{\alpha_{KA}\}=(\psi_{\nu,l_1},\psi_{\nu,l_1+1},\ldots,\psi_{\nu,l_m},\psi_{\nu,l_m+1})\,,
\end{equation}
composed from  $m$ pairs of neighbour physical states.
Krein-Adler theorem \cite{Krein,Adler} guarantees that the resulting  system 
described by the Hamiltonian operator of the form
\begin{eqnarray}
\label{deformed1}
&\mathcal{H}^{KA}_{(\nu,m)}=\mathcal{H}_{\nu+m}+4m+\frac{F_\nu(x)}{Q_\nu(x)}&
\end{eqnarray}
is  nonsingular on $\R^+$. Here
$F_\nu(x)$ and $Q_\nu(x)$ are real-valued  polynomials,   
$Q_\nu(x)$ has no zeroes on $\R^+$, its degree  is two more than that  
of $F_\nu(x)$, and so, the last rational term in (\ref{deformed1}) vanishes at infinity. 
The spectrum of the system (\ref{deformed1}) 
is the equidistant spectrum of the AFF model with  the removed energy levels
corresponding to the seed states. Consequently,  any gap 
in the resulting system has a size $12+8k$, where $k=0,1,\ldots$ 
corresponds to  $k$  adjacent pairs in the set (\ref{alpKA}) 
which produce a given gap.
An example of this kind of the systems is 
generated by the scheme $(\psi_{\nu,2},\psi_{\nu,3})$,
whose dual negative scheme is given by 
equation (\ref{Adlerkreineg}). 
\vskip0.1cm

Another class of rationally extended AFF systems 
is provided by  isospectral deformations generated by the
schemes  of  the form 
\begin{equation}\label{isoscheme}
\{\alpha_{iso}\}=(\psi_{\nu,-s_1},\ldots,\psi_{\nu,-s_m})\,,
\end{equation}
which contain only Wick rotated  states 
 $\rho_2(\psi_{\nu,n}(x))=\psi_{\nu,n}(ix)$.
As the functions used in this scheme are proportional to  $x^{\nu+1}$ 
and do not have real zeros other than $x=0$,  one obtains a regular 
on $\R^+$ system of the form
\begin{eqnarray}
\label{deformed2}
&\mathcal{H}^{iso}_{(\nu,m)}=\mathcal{H}_{\nu+m}+2m+f_\nu(x)\,,&
\end{eqnarray}
where $f_\nu(x)$ is a rational function disappearing at infinity  \cite{Grand},
and one can find that  potential of the system (\ref{deformed2}) is a convex on $\R^+$ function.
In this case the transformation does not remove or add energy levels, 
and,  consequently, the initial system $\mathcal{H}_\nu$  and the deformed system (\ref{deformed2}) 
are completely isospectral super-partners.
 Some concrete examples of the systems  (\ref{deformed2}) with 
 integer values of $\nu$ were  considered in \cite{CIP}. 
 \vskip0.1cm
 Consider yet another generalized Darboux scheme 
 which allows us to interpolate between different
 rationally deformed AFF systems.
 For this we assume that the initial AFF system
 is characterized by the parameter
 $\nu=\mu+m$, where 
 $-1/2<\mu\leq 1/2$ 
and $m$ can take any non-negative integer value. 
For these ranges of values of the parameter $\nu$,
real zeros of the functions $\psi_{\mu+m,n-m}$ are located between 
zeros of $\psi_{-(\mu+m)-1,n}$,
so that we can rethink the Krein-Adler theorem
 and consider the scheme 
\be
\label{Interpol}
\{\gamma_{\mu}\}=(\psi_{-(\mu+m)-1,n_1},\psi_{(\mu+m),n_1-m},\ldots,
\psi_{-(\mu+m)-1,n_{N}},\psi_{(\mu+m),n_{N}-m})\,,
\ee
which includes  $2N$ states and  where 
we suppose that $n_i-m\geq 0$
for all $i=1,\ldots, N$. 
The Darboux transformation based on the set (\ref{Interpol}) produces 
the system 
\be
\label{Intersys}
\mathcal{H}_{\mu+m}^{def}:=\mathcal{H}_{\mu+m}-2(\ln W(\gamma_\nu))''=\mathcal{H}_{\mu+m}+4N+h_{\mu+m}(x)/q_{\mu+m}(x)\,,
\ee
where the term  $4N$ is provided by the Gaussian factor in the Wronskian, 
and the last term is a rational function vanishing at infinity 
and having  no zeros in the whole real line, including the origin, if an only if $-1/2<\mu\leq1/2$,
 see Appendix \ref{apWron}. 
 Let us analyze now some special values of $\mu$
 
\textit{The case $\mu=0$}\,: by virtue of  relation between Laguerre and Hermite polynomials, 
$H_{2n}(x)=(-4)^nn!L_{n}^{(-1/2)}(x^2)$ and 
$H_{2n+1}(x)=2(-4)^nn!xL_{n}^{(1/2)}(x^2)$, in this case we obtain 
those systems which were generated in \cite{CIP} by Darboux transformations of the half-harmonic
oscillator. They are characterized  by gaps of the size $8+4k$, and represent 
 rational extensions of the  AFF model with integer  coupling constant $g=m(m+1)$, 
which in the case of $m=0$ reduce to  a rationally extended harmonic oscillator supplied with 
a potential barrier at $x=0$. Note that the minimal size of the gaps here is less than  
that for the systems produced by the scheme (\ref{alpKA}). 

\textit{The case $\mu=1/2$}\,:  we have here the relation    
$\rho_1(\psi_{m+1/2,n_i})=\psi_{-m-3/2,n_i}=(-1)^{m+1}\psi_{m+1/2,n_i-m-1}$,
 due to which 
the scheme (\ref{Interpol})
transforms into 
\be
\{\gamma_{{1}/{2}}\}=(\psi_{1/2+m,n_1-m-1},\psi_{1/2+m,n_1-m},\ldots,
\psi_{1/2+m,n_{N}-m-1},\psi_{1/2+m,n_{N}-m})\,,
\ee 
which corresponds to  (\ref{alpKA}) with
$l_i=n_i-m-1$. We additionally suppose 
that $n_i-m-1\not=n_{i-1}-m$, otherwise the Wronskian vanishes. Note that 
when $\mu\not=1/2$, the image of the states $\psi_{\mu+m,n_i-m-1}$
under Darboux mapping (\ref{Darstates})
is a physical state, but in the case $\mu=1/2$ such states are mapped into zero 
since  the argument  $\psi_{1/2+m,n_i-m-1}$  appears twice 
in the Wronskian of the numerator.

\textit{The case $\mu=-1/2$}\,: this case was not included
 in the range of $\mu$ from the beginning due to relation
$\rho_1(\psi_{m-1/2,n_i})=\psi_{-m-1/2,n_i}=(-1)^{m}\psi_{m-1/2,n_i-m}$
which 
would mean  the appearance of the repeated states in the scheme 
(\ref{Interpol}) and vanishing of the corresponding Wronskian.
However, in Appendix \ref{apWron} we show that  the  limit  relation 
$\lim_{\mu\to-1/2}{W(\{\gamma_\mu\})}/{(\mu+\frac{1}{2})^N}\propto
W(\{\gamma\})$ is valid, 
where the scheme $\{\gamma\}$ is
\be
\label{Interpol2}
\{\gamma\}=(\psi_{m-1/2,n_1-m},\Omega_{m-1/2,n_1-m},\ldots,\psi_{m-1/2,n_N-m},\Omega_{m-1/2,n_N-m})\,,
\ee
which 
corresponds to a non-singular  confluent Darboux transformation 
\cite{confDarb}. 

By considering this last comment, in conclusion we have that 
when $-1/2\leq\mu<1/2$, the states $\psi_{-(\mu+m)-1,n_i}$ 
(and $\Omega_{m-1/2,n_i-m}$ in the case of $\mu=-1/2$) are non-physical states.
This  means that 
 only the physical states 
$\psi_{\nu+m,n_i-m}$ indicate the energy levels removed under
 the corresponding Darboux transformation, i.e., there are gaps 
of the minimum size $2\Delta E=8$, where $\Delta E=4$ is the 
distance between energy levels of the AFF model, which can merge 
to produce energy gaps of the size $8+4k$.
On the other hand, when $\mu=1/2$,
we have a typical Krein-Adler  scheme with gaps of the size $12+4k$.

\begin{figure}[H]
\begin{center}
\includegraphics[scale=0.8]{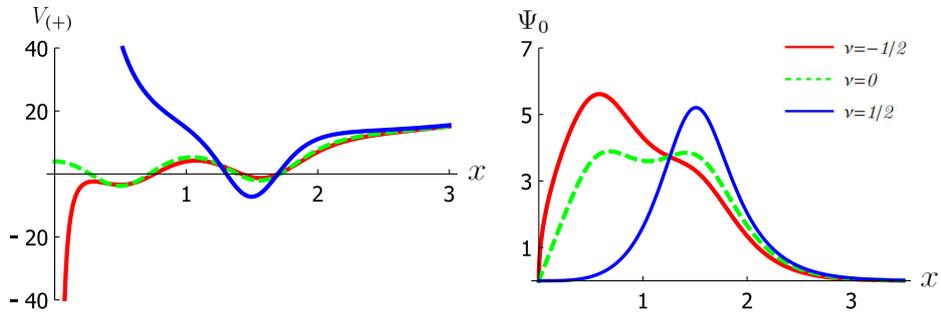} 
\caption{\small{On the left, 
a graph of the corresponding potential is shown which is 
produced by the 
associated Darboux transformation applied to the AFF model
with three indicated values of the parameter $\nu$ 
versus  the dimensionless coordinate $x$.
For $\nu=-1/2$,  
the corresponding limit is taken, and the resulting system has an attractive potential with a (not shown) 
potential 
barrier at $x=0$.
For $\nu=0$, we obtain a rationally extended half-harmonic 
oscillator. The case $\nu=1/2$ corresponds to the
Krein-Adler scheme $(\psi_{1/2,1},\psi_{1/2,2})$ with a 
gap equal to $12$. On the right, 
the ground states 
of the corresponding generated systems are shown 
as functions of dimensionless coordinate $x$.}}
\label{potential1}
\end{center} 
\end{figure}

\begin{figure}[H]
\begin{center}
\includegraphics[scale=0.8]{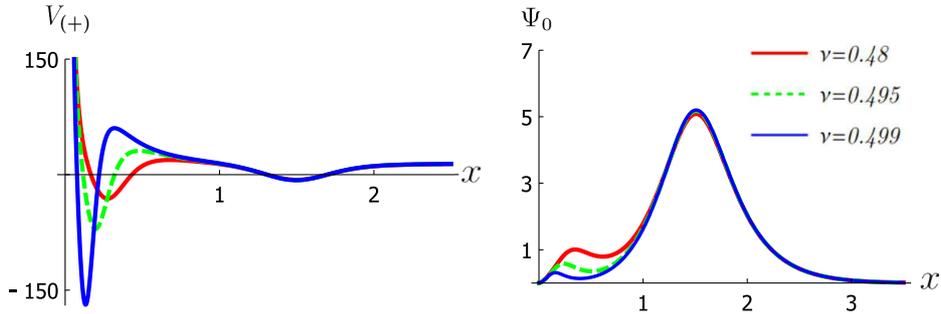} 
\caption{\small{On the left, the potential of deformed systems with $\nu$ close
 to $1/2$ is shown. On the right, the ground states of the corresponding 
 systems are displayed. }} 
\label{potential12}
\end{center} 
\end{figure}

 To give an example, we put $m=0$, that
  means $\nu=\mu$,  and consider the scheme $(\psi_{-\nu-1,2},\psi_{\nu,2})$
with $-1/2<\nu\leq1/2$,  whose Wronskian is presented explicitly 
 in Appendix \ref{Apendix-Wronskian}, and in the case of $\nu=-1/2$ we have 
the scheme $(\psi_{-1/2,2},\Omega_{-1/2,2})$.
 The potential of the rationally deformed AFF system 
 generated by the corresponding Darboux transformation
is shown in Fig. \ref{potential1}
and Fig. \ref{potential12}.

 As it is seen from the figures, 
 the first minimum of the potential grows in its absolute value, 
  its position moves to $0$,  and it disappears at $\nu=1/2$,
while  the  local maximum near zero also grows,
its position approaches zero,   and it goes to infinity  in the limit.
 Besides, the first maximum of the ground state 
 vanishes when $\nu$ approximates the limit value $1/2$. 
 Coherently with the described behavior of the potential, 
 the image of the  Darboux-transformed state $\psi_{\nu,1}$, 
 which is  the first excited state of the new system when $-1/2\leq \nu<1/2$,  vanishes when 
 $\nu\rightarrow 1/2$, the corresponding energy level disappears from the spectrum at $\nu=1/2$, 
 and the size of the gap increases from $8$ to $12$.
  
The described three possible selection rules to choose the
seed states correspond to  the negative scheme  (\ref{isoscheme}), which 
generates  isospectral deformations,  
the  positive Krein-Adler scheme (\ref{alpKA}),
 and the positive interpolating   scheme (\ref{Interpol}). 
Then we can apply the algorithm 
constructed in Section \ref{DualSec} to obtain 
the corresponding dual schemes for them. 
The positive and negative  dual schemes 
will be used in the next subsection
 to construct complete sets 
of the spectrum generating ladder 
operators for the rationally deformed 
conformal mechanics systems.

\section{Intertwining and ladder operators}\label{SecIntLad}

As a starting point, we consider  
any positive scheme for the 
AFF model $\mathcal{H}_\nu$ that produces its certain 
non-singular rational deformation. 
For simplicity we do not touch
here the schemes that contain Jordan states. However, 
we have relations (\ref{Polly2}) and  (\ref{spanChalf}), 
and relations (\ref{eqschemes3}) and (\ref{eqschemes4})
which
were extended to such cases with the corresponding substitutions; see the 
comments for Eq.  (\ref{DualSchemeJor1}). This means that 
the properties summarized below are also valid for the schemes containing Jordan states.
    
Let  a  positive scheme contains $n_+$ states labeled by  
$n_i$, $i=1,\ldots,n_+$, with  $n_{n_+}$ being the biggest 
quantum number in the set. We denote by $\mathcal{H}_{(+)}$ the system generated   
by the  Darboux transformation based on the set of the chosen seed states. 
By applying the algorithm from  Sec. \ref{DualSec}, we obtain
the corresponding dual negative scheme  
with $n_-=2n_{n_+}+2-n_+$ seed states
labeled by index $-l_j$ with  $j=1,\ldots,n_-$ and $-l_{n_-}=-n_{n_+}$.  
The resulting system of the  Darboux transformation 
based on  the negative scheme  we denote by
$\mathcal{H}_{(-)}$. By using
(\ref{eqschemes3}) we obtain that the generated Schr\"odinger operators 
are mutually shifted for a constant, 
\be
\label{dualL}
\mathcal{H}_{(+)}-\mathcal{H}_{(-)}=\Delta E(n_{n_+}+1)=2(n_++n_-)\,,\qquad \Delta E=4\,.
\ee
We can construct  the corresponding intertwining operators of both schemes 
by following the rule (\ref{generic-inter}). 
Let us  denote by  $A_{(+)}^\pm$ and 
$A_{(-)}^\pm$ the intertwining operators  of the positive and negative schemes
being differential operators of  the orders $n_+$ and $n_-$, respectively.
Some useful properties of these operators are summarized as follows.
First,  they satisfy  the intertwining relations 
\be
\label{inter-relation}
A_{(\pm)}^-\mathcal{H}_{\nu}=\mathcal{H}_{(\pm)}A_{(\pm)}^-\,,\qquad A_{(\pm)}^+\mathcal{H}_{(\pm)}=
\mathcal{H}_{\nu}A_{(\pm)}^+\,,
\ee
from where one concludes that the operators $A_{(\pm)}^-$ 
map differently physical eigenstates of  $\mathcal{H}_\nu$ 
as well as  non-physical ones obtained from them by action of
generators of the $K_4$ group. 
The  states 
$\widetilde{\psi}_{r(\nu),\pm n}$
behave asymptotically as  $e^{\pm x^2/2}$, 
and the states produced from them by application of
differential operators $A_{(\pm)}^-$   
will carry the same exponential factor. Having this asymptotic behavior in mind,
let us suppose that $\psi_{r(\nu),-l_*}$ and $\psi_{r(\nu),n_*}$ are some arbitrary states 
from the negative and positive scheme, respectively.   
By using  (\ref{inter-relation}), we obtain the relations
\begin{eqnarray}
\label{complement1}
&A_{(-)}^-\widetilde{\psi}_{r(\nu),-l_*}=A_{(+)}^-\rho_1(\psi_{r(\nu),n_{n+}-l_*})\,,\qquad
A_{(+)}^-\widetilde{\psi}_{r(\nu),n_*}=A_{(-)}^-\rho_1(\psi_{r(\nu),-(n_{n+}-n_*)})\,,\quad&
\end{eqnarray}
in both sides of which the functions satisfy the same 
second order differential equation and have the same behaviour at infinity. 
Note that in the dual schemes in 
(\ref{eqschemes3}) and (\ref{eqschemes4}),
the indexes $n_{n_+}-l_*$ and $-(n_{n_+}-n_*)$ are in correspondence with the indexes 
$r_i$, and $s_i$ of the states omitted from the positive and negative scheme, respectively.
 This helps us to 
obtain the identities 
\begin{equation}
\label{powerC}
A_{(-)}^+A_{(+)}^-=(-1)^{n_{n_+}+1-n_+}(\mathcal{C}_{\nu}^-)^{n_{n_+}+1}\,,
\qquad A_{(+)}^+A_{(-)}^-=(-1)^{n_{n_+}+1-n_+}(\mathcal{C}_{\nu}^+)^{n_{n_+}+1}\,.
\end{equation}
It is enough to prove the 
first relation in (\ref{powerC}), and the second is produced  by the 
Hermitian conjugation. As we know,  $A_{(+)}^-$, annihilates all  the 
states in the positive scheme, while  $A_{(-)}^+$ annihilates all 
the functions of the form $A_{(+)}^-\widetilde{\psi}_{r(\nu),-l_*}$. 
Then, acting by $A_{(-)}^+$ from the left  on both sides of the  first 
relation in (\ref{complement1}), 
we find  that 
 $\ker \big(A_{(-)}^+A_{(+)}^-\big)=(\psi_{\nu,0},\psi_{-\nu-1,0},\ldots,
\psi_{\nu,n_{n_+}},\psi_{-\nu-1,n_{n_+}})=\ker\, (\mathcal{C}_{\nu}^-)^{n_{n_+}+1}$.

Finally, to have a complete picture we write   
the relations
\begin{eqnarray}
\label{complement3}
A_{(-)}^-\psi_{r(\nu),k}=A_{(+)}^-\psi_{r(\nu),n_{n+}+1+k'}\,,\qquad 
A_{(+)}^-\psi_{r(\nu),-k'}=A_{(-)}^-\psi_{r(\nu),-(n_{n+}+1+k')}\,.&
\end{eqnarray}  
In the 
case of the dual schemes $\{\gamma\}$ and  
$\{\Delta_{-}^{(m-\frac{1}{2})}\}$ defined 
in (\ref{Interpol2}) and (\ref{DualSchemeJor1}), where $\nu=m-1/2$, similar relations 
are obtained but with $\psi_{-\mu-m-1,\pm n_i}$ and $\widetilde{\psi}_{-\mu-m-1,\pm n_i}$ replaced 
by $\Omega_{m-\frac{1}{2},\pm(n_i-m)}$ and 
$\breve{\Omega}_{m-\frac{1}{2},\pm(n_i-m)}$ when required.

With the help of the described  intertwining operators, we can construct 
three  types  of  ladder operators 
for $\mathcal{H}_{(\pm)}$ which are given by:
\begin{eqnarray}
\label{ladders}
&\mathcal{A}^{\pm}=A_{(-)}^-\mathcal{C}_{\nu}^\pm A_{(-)}^+\,, \quad
\mathcal{B}^{\pm}=A_{(+)}^-\mathcal{C}_{\nu}^\pm A_{(+)}^+\,,\quad
\mathcal{C}^{+}=A_{(-)}^-A_{(+)}^+\,,\quad \mathcal{C}^{-}=A_{(+)}^-A_{(-)}^+\,.&
\end{eqnarray}
Let us denote these operators in the compact form
$\mathcal{F}_a^{\pm}=(\mathcal{A}^\pm,
\mathcal{B}^\pm,\mathcal{C}^\pm)$,  $a=1,2,3$, and use 
(\ref{dualL}) and (\ref{inter-relation}) to obtain the commutation 
relations 
\begin{eqnarray}
\label{defsl2R}
&[\mathcal{H}_{(\pm)},\mathcal{F}_a^{\pm}]=\pm R_a\mathcal{F}_a^{\pm}\,,\qquad
[\mathcal{F}_{a}^-,\mathcal{F}_a^{+}]=\mathcal{P}_a(\mathcal{H}_{(\pm)})\,,&\\\nonumber&
\begin{array}{ll}
R_1=R_2=4\,,&\mathcal{P}_1=(\eta+2\nu+3)(\eta-2\nu+1)P_{n_-}(\eta)P_{n_-}(\eta+4)|_{\eta=\mathcal{H}_{(-)}-4}^{\eta=\mathcal{H}_{(-)}}\,,\\
&\mathcal{P}_2=(\eta+2\nu+3)(\eta+2\nu+1)P_{n_+}(\eta)P_{n_+}(\eta+4)|_{\eta=\mathcal{H}_{(+)}-4}^{\eta=\mathcal{H}_{(+)}}\,,\\
R_3=4(n_{n_+}+1)\,,&\mathcal{P}_3=P_{n_+}(\eta)P_{n_-}(\eta)|_{\eta=\mathcal{H}_{(+)}-4}^{\eta=\mathcal{H}_{(-)}}\,,
\end{array}&
\end{eqnarray}
where
\begin{eqnarray}
\label{Poly2}
P_{n_-}(y)=\prod_{i=1}^{n_-}(y-\lambda_{i}^-)\,,\qquad
P_{n_+}(y)=\prod_{i=1}^{n_+}(y-\lambda_{i}^+)\,,
\end{eqnarray} 
and  $\lambda_{i}^\pm$ are the corresponding eigenvalues 
of the seed states in the positive and negative schemes.
Equations (\ref{defsl2R}) are three different but related  copies of
the nonlinearly deformed conformal algebra 
$\mathfrak{sl}(2,\R)$. 
One can verify that due to the non-linearity of 
these three copies, the commutator of generators with different values of 
index $a$ do not vanish, and therefore 
the complete structure is rather complicated. 

Similarly to the non-deformed case, be means of a unitary transformation 
produced by $U=e^{-it\mathcal{H}_{(\pm)}}$ we obtain  the integrals of motion 
$\mathcal{F}_a^\pm(t)=e^{\mp R_a}\mathcal{F}_a$, and by linear combinations of them 
 construct the Hermitian generators 
$\mathfrak{D}_a(t)=(\mathcal{F}_a^-(t)-\mathcal{F}_a^+(t))/(i2R_a)$ and $
\mathfrak{K}_a(t)=(\mathcal{F}_a^+(t)+\mathcal{F}_a^-(t)+2\mathcal{H}_{(\pm)})/R_a^2$
which generate three copies of a
 non-linear deformation of the Newton-Hooke algebra, 
\begin{eqnarray}
[\mathcal{H}_{(\pm)},\mathfrak{D}_a]=-i\left(\mathcal{H}_{(\pm)}-\frac{(R_a)^2}{2}\mathfrak{K}_a\right)\,,\qquad
[\mathcal{H}_{(\pm)},\mathfrak{K}_a]=-2i\mathfrak{D}_a\,,\\\nonumber
[\mathfrak{D}_a,\mathfrak{K}_a]=\frac{1}{iR_a^3}\left(\mathcal{P}_a(\mathcal{H}_{(\pm)})-
2R_a\mathcal{H}_{(\pm)}+R_a^3\mathfrak{K}_a\right)\,,
\end{eqnarray}
which are hidden symmetries of the system described by $\mathcal{H}_{(\pm)}$.
 
In isospectral case, the operators $\mathcal{A}^\pm$ are the spectrum generating
ladder operators, where   
their action on physical eigenstates of $\mathcal{H}_{(\pm)}$ 
is similar  to that of  $\mathcal{C}_\nu^\pm$
in the AFF model. 
On the other hand, 
in rationally extended gapped systems obtained by   Darboux transfromations 
based on the  schemes  not containing  
Jordan states, the  separated  states 
have the form
$A_{(-)}^-\widetilde{\psi}_{-\nu-1,-l_j}=A_{(+)}^-\psi_{\nu,n_{n+}-l_j}$, where
the states  $\psi_{-\nu-1,-l_j}$
belong to the  negative scheme and $\psi_{\nu,n_{n+}-l_j}$ are the omitted 
states in the corresponding dual positive scheme. 
Since by construction the separated states
 belong to  the kernel of $A_{(-)}^+$, 
the operators $\mathcal{A}^\pm$ and $\mathcal{C}^-$ 
will always annihilate all them.

The resulting total picture related to the action of the ladder operators can be  summarized as follows. 
Operators of the $\mathcal{A}^\pm$ type detect all the separated states  organized in  valence bands, 
while they  act
like ordinary ladder operators in the equidistant part of the spectrum.  The lowering operator  $\mathcal{B}^-$ 
annihilates the lowest state in each valence band, and the raising operator $\mathcal{B}^+$ annihilates there
 the highest states,
and $\mathcal{B}^\pm$ also act in an ordinary way  in the equidistant part. 
The operators $\mathcal{C}^\pm$ connect the separated 
part of the spectrum  with its equidistant part, and  the lowering 
operator  $\mathcal{C}^-$ annihilates all the separated states 
as well as some excited  states in the equidistant part according to the rule:
if there is no level in the spectrum  of energy $E_n-\Delta E(n_{n_+}+1)$, 
where $\Delta E=4$, 
then the corresponding physical eigenstate of energy $E_n$ is annihilated by it.
For the case of the confluent Darboux transformations produced 
on the base of the scheme 
(\ref{Interpol2}) and its dual one, the separated states  are 
$A_{(-)}^-\widetilde{\psi}_{m-\frac{1}{2},-l_*}=
A_{(+)}^-\psi_{m-\frac{1}{2},n_N-m-l_*}$, 
but the picture  related to the  action of the ladder operators is  the same.

\section{Application: Example}
\label{Examples}
In this section we will apply the machinery of the dual schemes and the construction 
of nonlinear deformations of the conformal algebra to  a non-trivial example 
 of rationally extended  system with gaps.
  Remember that if we take 
 $\nu=\mu+m$, we do the change 
 $\psi_{-(\mu+m)-1,\pm n}$ by 
$\Omega_{-(\mu+m)-1,\pm(n-m)}$ with $n>m$ when $\mu\rightarrow-1/2$ in each of the relations 
that we have in the following,  see Section \ref{Mirror}.

\begin{figure}[H]
\begin{center}
\includegraphics[scale=0.6]{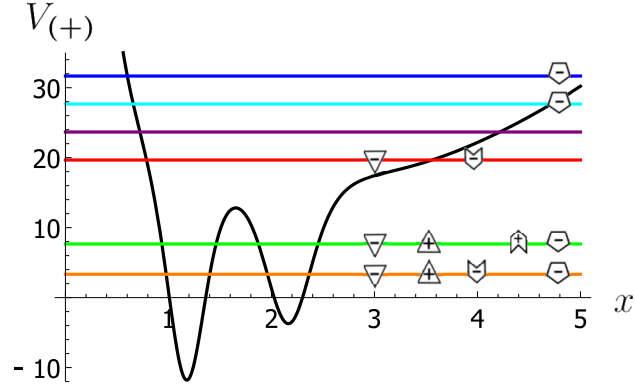} 
\caption{\small{The resulting potential with $\nu=1/3$ and energy levels of the system. 
The energy levels of the physical
states annihilated by the ladder operators $\mathcal{A}^-$, $\mathcal{A}^+$, $\mathcal{B}^-$, $\mathcal{B}^+$,
and  $\mathcal{C}^-$ are indicated from left to right.}} 
\label{Potential2}
\end{center} 
\end{figure}

Consider a system generated on the base of the Darboux-dual schemes 
$(\psi_{\nu,2},\psi_{\nu,3})\sim(\psi_{\nu,-0},
\psi_{\nu,-1},\psi_{\nu,-2},\psi_{-\nu-1,-2},\psi_{\nu,-3},\psi_{-\nu-1,-3})$. 
Here,   $n_-=2$, $n_+=6$,
  $n_{n_{+}}=n_{n_{-}}=3$ and $n_-+n_+=2(n_{n_+}+1)=8=
  2\Delta E$.
The positive scheme, whose Wronskian is
 given explicitly in Appendix \ref{Apendix-Wronskian}, corresponds to the Krein-Adler  scheme that    
provides us the system 
\begin{eqnarray}
\label{DeformedA-K}
&\mathcal{H}_{(+)}=-\frac{d^2}{dx^2}+V_{(+)}(x)\,,&
\end{eqnarray}  
whose potential $V_{(+)}$ is plotted in 
Figure \ref{Potential2}. The spectrum of the system, 
$\mathcal{E}_{\nu,0}=2\nu+3$, $\mathcal{E}_{\nu,1}=2\nu+7$,
$\mathcal{E}_{\nu,n}=2\nu+4(n+2)+3$, $n=2,\ldots$, 
is characterized by the presence of  the gap of the size $3\Delta E=12$, 
which appears between the first and second excited states. 
The negative scheme generates the shifted Hamiltonian  operator
$\mathcal{H}_{(-)}=\mathcal{H}_{(+)}-4\Delta E$.  
In terms of the intertwining operators $A_{(+)}^\pm$ and $A_{(-)}^\pm$
of the respective positive and negative schemes,  
the physical eigenstates of (\ref{DeformedA-K})
are given by
\begin{eqnarray}
\Psi_j&=&A_{(+)}^-\psi_{\nu,j}=A_{(-)}^-\widetilde{\psi}_{-\nu-1,j-3}\,,\qquad j=0,1\,,\\
\Psi_j&=&A_{(+)}^-\psi_{\nu,j+2}=A_{(-)}^-\psi_{\nu,j-2}\,,\qquad j=2,3,\ldots\,.
\end{eqnarray}

\noindent
The explicit form of the polynomials (\ref{Poly2}) for the system is
\begin{eqnarray}\label{Pn+y}
P_{n_+}(\eta)=(\eta-11-2\nu)(\eta-15-2\nu)\,,\\
P_{n_-}(\eta)=(\eta+9-2\nu)(\eta+13-2\nu)\prod_{i=0}^{3}(\eta+4n+3+2\nu)\,,\label{Pn-y}
\end{eqnarray}
and so,  $A_{(\pm)}^-A_{(\pm)}^+=P_{n_\pm}(\mathcal{H}_{\nu})$ and $A_{(\pm)}^-A_{(\pm)}^+=P_{n_\pm}(\mathcal{H}_{(\pm)})$.

The spectrum generating ladder operators are given by Eq.  (\ref{ladders}), 
and the nonlinearly deformed  conformal  algebras generated by each corresponding 
pair of the ladder operators and the Hamiltonian $\mathcal{H}_{(+)}$  are obtained from 
(\ref{defsl2R})  by using polynomials (\ref{Pn+y}) and (\ref{Pn-y}). 
To clarify  physical nature of the ladder operators in more detail, 
we inspect their corresponding kernels by using  relations (\ref{tools2}) and (\ref{powerC}).

We consider first the operators  $\mathcal{B}^\pm$, which have the  lowest differential
order $6$
 and possess  the following kernels\,: 
\begin{eqnarray}\nonumber
\ker \mathcal{B}^-=\text{span}\{A_{(-)}^-\widetilde{\psi}_{-\nu-1,-0}, 
A_{(+)}^-\widetilde{\psi}_{\nu,2},A_{(+)}^-\widetilde{\psi}_{\nu,3},
\Psi_{0},A_{(+)}^-\psi_{-\nu-1,0},\Psi_{2}\}\,,\\
\ker \mathcal{B}^+=\text{span}\{ A_{(+)}^-\widetilde{\psi}_{\nu,2},A_{(+)}^-\widetilde{\psi}_{\nu,3},
A_{(+)}^-\psi_{\nu,-0},A_{(+)}^-\psi_{-\nu-1,0},\Psi_{1},A_{(-)}^-\widetilde{\psi}_{-\nu-1,-1}\}\,,\nonumber
\end{eqnarray}
where only $\Psi_j$ with $j=0,1,2$ are physical states. 
The operators $\mathcal{B}^\pm$ act like fermionic operators 
in the separated two-dimensional valence band, in which 
$\mathcal{B}^+$ transforms  the state $\Psi_0$
into the excited state $\Psi_1$, and annihilates $\Psi_1$,
while $\mathcal{B}^-$ acts in a  similar  way but in the downward direction.

The ladder operators $\mathcal{A}^\pm$  have differential order $14$ 
and their  kernels  are
\begin{eqnarray}
\nonumber
&\ker \mathcal{A}^-=\text{span}\{ A_{(-)}^-\widetilde{\psi}_{\nu,-0},A_{(-)}^-\widetilde{\psi}_{\nu,-1},
A_{(-)}^-\widetilde{\psi}_{\nu,-2},A_{(-)}^-\widetilde{\psi}_{\nu,-3},\Psi_0,\Psi_1,\Psi_2, A_{(-)}^-\psi_{-\nu-1,0},&\\
&A_{(-)}^-\widetilde{\psi}_{-\nu-1,0},A_{(-)}^-\psi_{-\nu-1,-1}, 
A_{(+)}^-\widetilde{\psi}_{\nu,1},
A_{(+)}^-\widetilde{\psi}_{-\nu-1,1},
A_{(+)}^-\widetilde{\psi}_{-\nu-1,2},
A_{(+)}^-\widetilde{\psi}_{-\nu-1,3}\}\,,\nonumber&\\
\nonumber
&\ker \mathcal{A}^+=
\text{span}
\{ A_{(-)}^-\widetilde{\psi}_{\nu,-0},
A_{(-)}^-\widetilde{\psi}_{\nu,-1},
A_{(-)}^-\widetilde{\psi}_{\nu,-2},
A_{(-)}^-\widetilde{\psi}_{\nu,-3},\Psi_0,\Psi_1,
A_{(-)}^-\widetilde{\psi}_{\nu,4},&\\
&
A_{(-)}^-\psi_{-\nu-1,-0},A_{(-)}^-\psi_{\nu,-4},
A_{(-)}^-\psi_{-\nu-1,-4}, 
A_{(-)}^-\widetilde{\psi}_{\nu,0},
A_{(-)}^-\widetilde{\psi}_{-\nu-1,0},
A_{(-)}^-\widetilde{\psi}_{-\nu-1,1},
A_{(-)}^-\widetilde{\psi}_{-\nu-1,2}\}.\nonumber&
\end{eqnarray}
Both  separated states are detected by $\mathcal{A}^\pm$: they are annihilated by both lowering and raising 
ladder operators. From the described  action of the ladder operators $\mathcal{A}^\pm$ and $\mathcal{B}^\pm$ it is clear 
that one cannot connect with their help the two separated states with the states in the equidistant part
of the spectrum, 
and we need another pair of the ladder operators to do this job. 
Fortunately, we have the operators 
$\mathcal{C}^\pm$ of differential order $8$, whose kernels are 
\begin{eqnarray}
&\ker \mathcal{C}^-=\text{span}\{ A_{(-)}^-\widetilde{\psi}_{\nu,-0},A_{(-)}^-\widetilde{\psi}_{\nu,-1},
A_{(-)}^-\widetilde{\psi}_{\nu,-2},A_{(-)}^-\widetilde{\psi}_{\nu,-3},\Psi_0,\Psi_1,\Psi_4,\Psi_5\}\,,&\nonumber\\
&\ker \mathcal{C}^+=\text{span}\{A_{(+)}^-\widetilde{\psi}_{\nu,2},A_{(+)}^-\widetilde{\psi}_{\nu,3},
A_{(+)}^-\psi_{\nu,-0},
A_{(+)}^-\psi_{\nu,-1},
A_{(+)}^-\psi_{\nu,-2}\,&\nonumber\\
&A_{(+)}^-\psi_{-\nu-1,-2},
A_{(+)}^-\psi_{\nu,-3},
A_{(+)}^-\psi_{-\nu-1,-3}, \}\,.&\nonumber
\end{eqnarray}
These equations show that the lowering operator $\mathcal{C}^-$
annihilates all the states in the separated valence band as well as some states in
the equidistant part of the spectrum, but the raising operator $\mathcal{C}^+$ does not annihilate 
any physical state.
In fact, by using the commutation relation for these operators given in (\ref{ladders}), 
which for this case is read as $[\mathcal{H}_{(+)},\mathcal{C}^{\pm}]=\pm 4\Delta E\mathcal{C}^{\pm}$,
one can derive  that $\mathcal{C}^+ \Psi_{0}=\Psi_{2}$ and   $\mathcal{C}^+ \Psi_{1}=\Psi_{3}$.
This allows us  to connect the separated states with the equidistant part of the spectrum. 

\section{Summary, discussion and outlook}
\label{Discussion}

We studied the effect of the Klein four-group $K_4$ 
as a symmetry of the time-dependent Schr\"odinger equation 
of the AFF model, and found that it has deep implications in dependence on
the values of the parameter $\nu$
 in the coupling constant $g(\nu)=\nu(\nu+1)\geq-1/4$. 
In general, the action of the $K_4$ transformations 
 changes the values of the energy, and transforms 
physical eigenstates into non-physical ones. In the case 
$\nu=\ell-1/2$ with $\ell=0,1,2,\ldots$, however, 
the reflection symmetry of the coupling constant, $\rho_1:\nu\rightarrow-\nu-1$, 
reduces to the identity transformation when $\ell=0$, while  for $\ell\geq 1$ it 
annihilates the first $\ell$ states and transforms all other eigenstates,
$\rho_1(\psi_{\ell-1/2,n+l})=(-1)^{\ell}\psi_{\ell-1/2,n}$,
coherently with lowering their energies
so that the spectrum is not changed.  
Thus, the $\rho_1$ appears to be a true, nontrivial  $\Z_2$ 
symmetry of the system when
the parameter $\nu$
takes half-integer values.
Omitting the normalization constants in wave functions $\psi_{\ell-1/2,n}$,
$n=0,\ldots,\ell-1$,  $\ell\geq 1$, annihilated by
$\rho_1$, the same transformation  allows us to construct 
non-physical states which play an important role in our 
constructions and can be transformed into physical states by 
the action of the raising operator. On the other hand, 
the spatial Wick rotation $\rho_2:(x,E_{\nu,n})\rightarrow (ix,-E_{\nu,n})$
corresponds to the automorphism of the conformal algebra $\mathfrak{sl}(2,\R)$.
This automorphism  transforms  an 
infinite-dimensional unitary representation of $\mathfrak{sl}(2,\R)$
restricted 
from below, which is realized on the spectrum of  the AFF system,
into an infinite-dimensional unitary representation restricted 
from above~\footnote{The restricted from above unitary representations 
 are non-physical from the 
point of view of 
the AFF system,  but  both types  of the 
$\mathfrak{sl}(2,\R)$-representations find 
applications in the theory of anyons \cite{PlyAnyons,JacNai,HPV}.}.
The non-physical states generated by the transformations
of the $K_4$ group appear in the kernels 
of the degrees of the decreasing and increasing generators of 
 $\mathfrak{sl}(2,\R)$.
In special cases of $\nu=\ell-1/2$, some of those 
non-physical states are changed for Jordan states
of the AFF system.

We showed that in the $\mathcal{N}=2$  super-extensions of the AFF model,
the non-physical eigenstates produced  by the $K_4$ 
generators play a key role in the corresponding 
Darboux transformations, and that  
the $K_4$ is the automorphism group   
of their superconformal $\mathfrak{osp}(2,2)$ dynamical symmetry. 
The interesting feature of the transformations 
generated by the $\rho_1$ is that they change the systems
with exact (unbroken) $\mathcal{N}=2$  Poincar\'e supersymmetry
into the systems in a  spontaneously broken phase, 
and vise-versa.  The  peculiarity of the case $\nu=-1/2$ 
reveals  itself here once again: both possible supersymmetric 
extensions of the AFF model
turn out to be related by a unitary transformation provided by $\rho_1$, and are
described by the unbroken $\mathcal{N}=2$  Poincar\'e supersymmetry.

Then we  use  the  discrete transformations generated by the Klein four-group
together with the conformal symmetry 
to generate,  by means of Darboux transformations, 
infinite families of
new, exactly  solvable quantum systems
with  
arbitrary number of gaps introduced into the 
equidistant spectrum of the AFF model.  
The obtained systems are described
by symmetries of the type of finite $W$ algebras, 
which represent nonlinearly deformed and extended
forms of the  conformal $\mathfrak{sl}(2,\R)$ algebra.

To construct such rational deformations of the AFF system
and study their spectral properties,
we
 developed the algorithm 
 of the dual Darboux schemes  
 for the  conformal mechanics  model with arbitrary values of the statistics parameter 
 $\nu$.  Note that the simplest form of the dual Darboux schemes 
 appears in the construction of the $\mathcal{N}=2$ super-extensions 
 of the AFF model and generators of 
 their superconformal $\mathfrak{osp}(2,2)$ symmetry.
 The physical eigenstates  together with non-physical ones  generated from them 
 by transformations of the Klein four-group  form a base
 for the Darboux duality in the case of  $\nu\neq \Z+1/2$. In the case of half-integer 
 values of $\nu$ the Jordan states naturally enter the construction via the 
 confluent Darboux transformations. 
 Each pair of the dual Darboux transformations different from that 
 we employed in the construction of the  
 $\mathcal{N}=2$ super-extensions 
 of the AFF model,  generates its some rational 
 deformation.  The obtained in such a way system  can be completely isospectral 
 (up to a global spectral shift) to
 the initial conformal mechanics  model, or may have  a finite number of valence bands in the low 
 part of the spectrum, which are separated by gaps 
 between themselves and from the semi-infinite band with equidistant energy levels.
 The minimal size of a  gap in our construction corresponds to one missing energy level
 in comparison with two missing levels in gaps of minimal size in the systems generated by the
 Krein-Adler transfromations, which  also are  included in our dual Darboux 
 schemes.
 
 We showed that when the statistics parameter varies continuously,
 the spectrum of rationally deformed AFF systems 
 suffers structural changes  at  half-integer (``fermionic")
  values of  $\nu$.
No such changes happen, however,  at integer values of $\nu$
corresponding to the case of bosons in the context of 
the statistics transmutations \cite{LeiMyr,MKWil,Poly}.
Recall  that all the deformations of the conformal mechanics model
 with $\nu\in\Z$
can be generated by generalized Darboux transformations 
from the quantum harmonic oscillator system \cite{CIP}.
At the same time we also note here that
the question of the self-adjoint extension of the AFF 
Hamiltonian operator requires a special consideration
in the case $\nu=-1/2$ \cite{KirLoy}, which corresponds 
to a minimal value of the coupling constant $g$
for which the spectrum is bounded from below,
and when, as we saw, the Klein four-group 
symmetry suffers minimal breaking.

 The Darboux duality  allowed us to obtain the set of the three pairs 
of ladder operators of different but complementary nature.  
In the case of  the  rationally deformed gapped systems, 
Hermitian conjugate ladder operators of the $\mathcal{A}$ type detect all the separated states,
each of which is annihilated by both, the lowering   and  the raising, 
operators;  the lowering operator $\mathcal{A}^-$  also annihilates the lowest state in the 
equidistant part of the spectrum. 
The raising ladder operator of the  $\mathcal{B}$ type detects the states with highest energy level
in each valence band by annihilating them. The lowering operator  $\mathcal{B}^-$
makes the same  with the 
states of the lowest energy level in each valence band,  and also annihilates the lowest state
in the equidistant part of the spectrum. Although the operators of these two types detect
all the separated states as well as identify the borders of the valence bands 
 and the edge of the semi-infinite band with  equidistant energy levels, they  cannot
connect  the states from different bands. This job is realized 
with the help of  the ladder operators  of the $\mathcal{C}$ type.
As a result,  one can see that any of the two sets of the ladder operators, ($\mathcal{C}^\pm$, $\mathcal{A}^\pm$)
or ($\mathcal{C}^\pm$, $\mathcal{B}^\pm$), forms a complete set of the spectrum generating 
ladder operators
by which any eigenstate of the rationally deformed AFF system can be transformed 
into its any other eigenstate.
In the case if we have an isospectral 
deformation of the AFF system $\mathcal{H}_\nu$ obtained via the
Darboux scheme (\ref{isoscheme}), 
the operators  $\mathcal{A}^\pm$ are enough to generate
the entire tower of physical eigenstates starting from 
any physical eigenstate.

Each of the three pairs of the conjugate ladder operators 
 together with the Hamiltonian operator generate
 some nonlinearly deformed version of the conformal 
 $\mathfrak{sl}(2,\R)$  algebra  of the $W$-type \cite{Walg}, 
  which is the symmetry of the 
 corresponding rationally deformed AFF system of a  generic form.
We, however, did not compute commutators between 
ladder operators of different types, 
but with a quick inspection one can notice that new 
structures are generated.
  Though the 
 resulting picture is expected to be  rather complicated and requires a separate
 study, it should  be  similar to that appearing 
  in the case of $\nu=0$, which was analyzed in detail in \cite{LM2},
  as well as to that in the $PT$-regularized 
 two-particle Calogero systems \cite{JM2,PlyNon}, 
  and can be described  as follows. 
Any extended system
  composed from a pair of the AFF systems characterized by 
  the parameters 
  $\nu$ and $\nu+m$, $m\in\Z$, 
  are described, as we showed,  by the superconformal 
  $\mathfrak{osp}(2,2)$ symmetry in the case of $m=1$, 
  while a non-linear 
  deformation of this superalgebra should appear when $m>0$.  
  On the other hand, 
  if the composed system contains a pair of
  rationally deformed AFF systems, according to 
  our results in \cite{LM2}, one can expect  that 
   its 
  spectrum should be described by some non-linear 
  extensions of the
  $\mathfrak{osp}(2|2)$ symmetry.
  Some nonlinearly extended versions of  $\mathfrak{sl}(2,\R)$
  are expected then to appear as $W$-type algebras
  describing symmetries of the rationally deformed AFF systems.

 Our consideration of rational deformations of the 
  conformal mechanics was restricted by inclusion of  
  Jordan states of the simplest form.  
  Following the analysis and ideas presented in refs. 
  \cite{Confluent3,confDarb,Confluent4,Confluent1,Confluent2}, the  constructions can be  generalized
  to the case of higher order Jordan states defined via relations 
\begin{eqnarray}
(L-\lambda_*)\Omega_{*}^{(0)}=\psi_{*}\,,\qquad
(L-\lambda_*)\Omega_{*}^{(k)}=\Omega_{*}^{(k-1)}\,,\qquad k=1,\ldots\,,
\end{eqnarray}
as well as to their further generalizations
defined as the states annihilated by polynomial in $L$ operators 
 \cite{CarPly}.
The states that we have used correspond to  $\Omega_{*}^{(0)}$. 
In this way, one  can  
produce the systems by means of confluent Darboux transformations
  which involve more Jordan states of these chains, 
and one can expect that the spectrum of the resulting systems will have a 
similar gapped structure. 
Then it would be interesting to study  such kind of the systems from the point of view 
of the spectrum generating ladder operators and the
extended nonlinear deformations 
of the (super)conformal algebra associated with them. 

It is  known that the conformal symmetry 
underlies the relation between the quantum free particle 
and harmonic oscillator systems \cite{Nied,DGH,Gonera}.
A similar picture also is valid for the 
two-particle Calogero system without confining potential term 
and  omitted center of mass degree of freedom, i.e. 
for the system (\ref{conformalaction}),
and the AFF model (\ref{mostgeneralH}) \cite{AFF,App1,Steu,Oka2}. 
The Calogero model and its deformations, in turn,  are intimately related
to the soliton solutions of the Korteweg-de Vries equation
and higher equations of its hierarchy \cite{KdV1,KdV2,KdV3,JM}.
It would be interesting to investigate the question of a possible relation
between rational deformations of the AFF model studied here
and solutions to the same hierarchy of completely integrable
systems described by partial differential equations.
At the same time,   the approach based on the Klein
four-group transformations employed here can  also 
be applied to  the two-particle Calogero model 
with arbitrary values of the statistics parameter $\nu$ but without 
the confining harmonic potential term.
In this way one could expect to generate 
new quantum solvable systems which may be 
related to the Korteweg-de Vries hierarchy.

\vskip0.2cm

\noindent {\large{\bf Acknowledgements} } 
\vskip0.1cm

The work was partially supported by 
the CONICYT scholarship 21170053 (LI),
FONDECYT Project 1190842 (MSP),
and the Project USA 1899 (MSP and LI).

\appendix 
\section*{Appendix}

\section{Dual schemes in half-integer case}
\label{dualgamma}
To obtain the dual schemes in the half-integer case, 
we analyze first the relations
that exist  
between of $\mathcal{H}_{-1/2}$ and $\mathcal{H}_{-1/2+\ell}$.
The latter are given by the dual schemes  
$(\psi_{-1/2,\pm0},\ldots,\psi_{-1/2,\pm(\ell-1)})$,
whose Wronskians are  
\begin{eqnarray}
\label{Wrondual}
&W(\psi_{-1/2,\pm0},\ldots,\psi_{-1/2,\pm(\ell-1)})=x^{\ell^2/2}e^{\mp \ell x^2/2}\,.&
\end{eqnarray}
The corresponding intertwiners map eigen- and Jordan states of 
$\mathcal{H}_{-1/2}$ to those of $\mathcal{H}_{-1/2+\ell}$.
If we choose the scheme with positive indexes, some of these mappings
useful for the following are given by 
\begin{eqnarray}
\label{interLnu1}
\A_\ell^-\psi_{-1/2,n}=\psi_{-1/2+\ell,n-\ell}\,, \qquad 
\A_\ell^-\Omega_{\nu,-1/2}=\Omega_{-1/2+\ell,n-\ell}\,, \qquad n\geq\ell\,,\\
\A_\ell^-\Omega_{-1/2,l}=\psi_{-(-1/2+\ell)-1,l}\,,\qquad\l<\ell\,,
\end{eqnarray}
where $\A_\ell^-$  and its  Hermitian conjugate $\A_\ell^+$ 
are the  intertwining operators of the chosen Darboux transformation. 
On the other hand if we take the scheme with negative sign in indices, we obtain
another intertwining operators $\B_{\ell}^\pm$, which satisfy 
the relation $\B_{\ell}^\pm=(i)^\ell\rho_2(\A_{\ell}^\pm)$, i.e, 
their action on eigenstates and Jordan states can be obtained
by application of $\rho_2$ to  the relations that correspond to the 
 action of $\A_{m}^\pm$.  

Now, to derive the dual schemes let us assume that we have a collection
 of non-repeated seed states of the form 
$(\psi_{-1/2,0},\ldots,\psi_{-1/2,\ell-1},\{\vartheta_{-1/2}\})$, 
where  $\{\vartheta_{-1/2}\}$ contains $N_1$ arbitrary 
physical states 
$\psi_{-1/2,k_i}$ with
 $k_i>\ell-1$ for  $i=1,\ldots,N_1$, and  $N_2$ arbitrary Jordan states of 
 the form $\Omega_{-1/2,l_j}$
 with $j=1,\ldots,N_1$. 
 In the same way as we did in Section \ref{Mirror}, 
 we define $n_N$ as the largest of the numbers  $n_{N_1}$ and $n_{N_2}$, 
and also we suppose for simplicity that the signs of both $k_i$ and $k_j$ 
are positive.
Then we use (\ref{id1}) and (\ref{Wrondual}) to  write 
 $W(\psi_{-1/2,0},\ldots,\psi_{1/2,\ell-1},\{ \vartheta_{-1/2}\})=
 x^{\ell^2/2}e^{-\ell x^2/2}W(\{ \A_\ell^-\vartheta_{-1/2}\})$.
  The next step is to use 
the extension of the dual schemes for $\nu=-1/2$, i.e,
we change each function of the form 
$\psi_{-\nu-1,n}$ by $\Omega_{-1/2,n}$ in equation 
(\ref{eqschemes3}),  and 
use it to rewrite this last Wronskian relation as 
\begin{eqnarray}
\label{SchemeJor1}
&
W(\A_\ell^-\{ \vartheta_{-1/2}\})=x^{-\ell^2/2}e^{-(n_N+1-\ell/2)x^2}W(\{\Delta^{(-1/2)}_-\})\,,
&
\end{eqnarray}
where $\Delta_{-}^{(-1/2)}$ is the dual scheme of 
$(\psi_{-1/2,0},\ldots,\psi_{-1/2,\ell-1},\{\vartheta_{-1/2}\})$ given by 
$\{\Delta_{-}^{(-1/2)}\}=(\psi_{-1/2,-0},\ldots,\psi_{-1/2,-(\ell-1)},\{\vartheta_{-1/2}^-\})\,,$
and 
\begin{eqnarray}
\{\vartheta_{-1/2}^-\}=(\psi_{-1/2,-\ell},\Omega_{-1/2,-0},\ldots,\check{\psi}_{-1/2,-s_j},
\check{\Omega}_{-1/2,-r_i},\ldots,\psi_{-1/2,-n_{N}},\Omega_{-1/2,-n_{N}})\,.
\end{eqnarray}
Here, as well as in the non-half-integer case, 
the marked functions $\check{\psi}_{-1/2,-s_j}$ and  
$\check{\Omega}_{-1/2,-r_i}$  indicate  the omitted states with $s_j=n_{N}-l_j$ 
and $r_i=n_{N}-k_i$.
In the last step, we use  Eqs. (\ref{id1}) and (\ref{Wrondual}) with the 
negative sign to write the equality $
W(\{\Delta_{-}^{(-1/2)}\})=x^{\ell^2/2}e^{\ell x^2/2}W(\mathbb{B}_\ell^-\{\vartheta_{-1/2}^-\})
$ and as analog of  (\ref{SchemeJor1}) we obtain 
\begin{eqnarray}
\label{DualSchemeJor1}
W(\A_\ell^-\{\vartheta_{-1/2}\})=e^{-(n'_N+1)x^2}W(\mathbb{B}_\ell^-\{\vartheta_{-1/2}^-\})\,,
\qquad n'_N=n_N-\ell\,.
\end{eqnarray}
This relation is the dual scheme equation for the case $\nu=\ell-1/2$. By means of  
(\ref{interLnu1}) and its analogs for $\B_\ell^-$ obtained by the application of $\rho_2$, 
we conclude that in the scheme of the left hand side of the equation 
there are $N_1$ physical states
of the form
$\A_\ell^-\psi_{-1/2,k_i}=\psi_{\ell-1/2,k_i-\ell}$, and a mixture of $N_2$ Jordan states
and formal states produced by $\rho_2$ 
distributed in the following way:
we have Jordan states 
$\A_\ell^-\Omega_{-1/2,l_i}=\psi_{\ell-1/2,l_i}$ when $l_i<\ell-1$, and
formal states 
$\A_\ell^-\Omega_{-1/2,l_i}=\psi_{\ell-1/2,l_i-\ell}$  when $l_i\geq \ell$.
The omitted  states in the scheme on the right hand side are  
$\B_{\ell}^-\check{\psi}_{-1/2,-s_j}=\check{\psi}_{-1/2+\ell,-(s_j-\ell)}$
and  
$\B_{\ell}^-\check{\Omega}_{-1/2,-r_j}=\check{\psi}_{-\ell-1/2,-r_j}$ 
($\B_{\ell}^-\check{\Omega}_{-1/2,-r_j}=\check{\psi}_{-\ell-1/2,-(r_j-\ell)}$ )
when $r_j\leq \ell-1$  ($r_j>\ell$). 
Note that the largest index in both sides of the equation 
 is now given by $n_N'=n_N-\ell$. 
In comparison with the non-half-integer case, 
this is the same result that we would obtain 
if we consider equation (\ref{eqschemes3}) in the non-half-integer case, 
and then formally change the states 
of the form $\psi_{-\nu-1,l_i}$ by $\Omega_{-\ell-1/2,l_i-\ell}$  when 
$l_i\geq\ell$ in the limit 
$\nu\rightarrow \ell-1/2$. 

Relation analogous to  (\ref{eqschemes4})
would be obtained if we start from the case $\nu=-1/2$ with a scheme composed from
the eigenstates and Jordan states produced by $\rho_2$, and then 
apply the same arguments employed for the case analyzed above.

\section{ Some Wronskian relations}
\label{apWron}
We show   here that the Wronskian  (\ref{Interpol}) 
takes non-zero values and that it reduces to  (\ref{Interpol2}) 
in the limit $\mu\to-1/2$. For this, consider first a generic system (\ref{Sch}) which has a set of the seed 
states $(\phi_1,\phi_2,\ldots,\phi_{2l-1},\phi_{2l})$ 
 with eigenvalues $\lambda_1<\lambda_2<\ldots<\lambda_{2l-1}<\lambda_{2l}$. Then 
 the following relation 
\begin{equation}
\label{producto1}
W(\phi_1,\phi_2,\ldots,\phi_{2l-1},\phi_{2l})=\prod_{i=0}^{l-1}W(\A_{2i}\phi_{2i+1},\A_{2i}\phi_{2i+2})\,,
\end{equation}
can be proved by induction, where $\A_{0}=1$, and $\A_{2i}$  with $i\geq 1$ corresponds to the intertwining 
operator associated with the scheme $(\phi_1,\ldots,\phi_{2i})$. 
{}From (\ref{producto1})
it follows that if  each factor 
$W(\A_{2i}\phi_{2i+1},\A_{2i}\phi_{2i+2})$ does not have zeros, 
then the complete  Wronskian neither has. 
To inspect the properties of the   Wronskian factors, we use the relation
\begin{equation}
\label{Wronskianderivative}
W'(\A_{2i}\phi_{2i+1},\A_{2i}\phi_{2i+2})=(\lambda_{2i+2}-\lambda_{2i+1})\A_{2i}\phi_{2i+1}\A_{2i}\phi_{2i+2}\,,
\end{equation} 
and integrate it from $a$ to $x$, 
\begin{equation}
\label{WronskianAphi}
W(\A_{2i}\phi_{2i+1},\A_{2i}\phi_{2i+2})=(\lambda_{2i+2}-\lambda_{2i+1})\int_a^x\A_{2i}
\phi_{2i+1}\A_{2i}\phi_{2i+2}d\zeta+\omega\,,
\end{equation}
where $ \omega=W(\A_{2i}\phi_{2i+1},\A_{2i}\phi_{2i+2})|_{x=a}$. In the case when  
functions $\A_{2i}\phi_{2i+1}$, $\A_{2i}\phi_{2i+2}$ and their first derivatives vanish 
in $b$,  we find  $\omega=-(\lambda_{2i+2}-\lambda_{2i+1})\int_a^b\A_{2i}\phi_{2i+1}\A_{2i}\phi_{2i+2}d\zeta$,
and then
\begin{equation}
\label{WronskianAphi2}
W(\A_{2i}\phi_{2i+1},\A_{2i}\phi_{2i+2})=-(\lambda_{2i+2}-\lambda_{2i+1})\int_x^b\A_{2i}\phi_{2i+1}
\A_{2i}\phi_{2i+2}d\zeta\,.
\end{equation}
Relation (\ref{producto1}) takes then the form
\begin{equation}
\label{Wronskianintegral}
W(\phi_1,\phi_2,\ldots,\phi_{2l-1},\phi_{2l})=\prod_{i=0}^{l-1}(\lambda_{2i+1}-\lambda_{2i+2})\int_x^b\A_{2i}\phi_{2i+1}\A_{2i}\phi_{2i+2}d\zeta_i\,.
\end{equation}

Analogously, one can consider a generic system,  choose $l$ solutions $\varphi_i$
of Eq. (\ref{Sch}), and construct $l$ corresponding Jordan states  $\Omega_i$ using Eq.
(\ref{omega1}). Assuming also that these states satisfy relations (\ref{omega2}), 
one can find that
\begin{equation}
\label{producto2}
W(\varphi_1,\Omega_1,\ldots,\varphi_{l},\Omega_l)=\prod_{i=0}^{l-1}W(\A_{2i}^{\Omega}\varphi_{i+1},\A_{2i}^{\Omega}\Omega_{i+1})=
\prod_{i=0}^{l-1}\int_{x}^{b}(\A_{2i}^{\Omega}\varphi_{i+1})^2d\zeta_i\,,
\end{equation}
where $
\A_{0}^{\Omega}=1$
and $\A_{2i}^{\Omega}$ correspond to the intertwining operator associated with the scheme
 $(\varphi_1,\Omega_1\ldots,\varphi_{l},\Omega_l)$.
 Relation  (\ref{producto2}) can be proved in a  way similar to that 
 for (\ref{Wronskianintegral}). 

Let us turn now to the AFF model, where $a=0$, $b=\infty$, 
and choose the seed states in (\ref{producto1}) in correspondence with our 
picture: 
for $i=0,\ldots,l-1$
we fix $\phi_{2i+1}=\psi_{-\mu-m-1,n_{i+1}}$ and $\phi_{2i+2}=\psi_{\mu+m,n_{i+1}-m}$. 
This identification implies that 
$\lambda_{2i+1}=E_{-\mu-m-1,n_{i+1}}$, $\lambda_{2i+2}=E_{\mu+m,n_{i+1}-m}$, and  
$\lambda_{2i+2}-\lambda_{2i+1}=4(\mu+1/2)$. These both functions
and their first derivatives  
behave for large values of  $x$  as $e^{-x^2/2}$, and  vanish at $x=\infty$. 
This behavior is not changed by application of any differential operator 
with which we work.
 On the other hand,  near zero we have 
 $\A_{2i}\psi_{-\mu+m+1,n_{i+1}}\sim
x^{-\mu-m-i}$ and $\A_{2i}\psi_{\mu+m,n_{i+1}-m}\sim
x^{\mu+m+1+i}$. Therefore, for small values of $x$, 
$\A_{2i}\psi_{-\mu+m+1,n_{i+1}}\A_{2i}\psi_{\mu+m,n_{i+1}-m}\sim x$,
and $W(\A_{2i}\psi_{-\mu+m+1,n_{i+1}},\A_{2i}\psi_{\mu+m,n_{i+1}-m})$ 
takes a finite value when $x\rightarrow 0^+$. 
Knowing this and  Eq. (\ref{Wronskianderivative}),   we employ
the Adler method \cite{Adler}, and use the theorem on nodes of 
wave functions  to show that 
zeros and the minima and maxima of the functions 
$\A_{2i}\psi_{-\mu+m+1,n_{i+1}}$ and $\A_{2i}\psi_{\mu+m,n_{i+1}-m}$
do not coincide,  and that their corresponding Wronskian 
is non-vanishing.

 In the case $\mu=-1/2$, 
we put $\varphi_{j}=\psi_{m-1/2,n_{j+1}-n}$ with $j=0,\ldots,l-1$,  and 
then we arrive at  the relations 
\begin{eqnarray}
\label{WronskianAphi3}
&\frac{W(\{\gamma_\mu\})}{(4\mu+2)^N}
=(-1)^{l}
\prod\limits_{i=0}^{l-1}\int_x^\infty\A_{2i}\psi_{-\mu-m-1,n_{i+1}}
\A_{2i}\psi_{\mu+m,n_{n_{i+1}}-m}d\zeta_i\,,&\\&
W(\{\gamma\})=\prod\limits_{j=0}^{l-1}\int_x^\infty(\A_{2j}^{\Omega}\psi_{m-1/2,n_{j+1}-m})^2d\zeta_j\,,&
\end{eqnarray}
where the sets $\{\gamma_\mu\}$ and $\{\gamma\}$
are defined in (\ref{Interpol}) and (\ref{Interpol2}). We note that 
both equations are pretty similar each other, and 
 if we suppose that $\A_{2i}\to\A_{2i}^{\Omega}$ when $\mu\to-1/2$, and 
 take into account  the relation
$\psi_{m-1/2,n_{j}-m}\propto\psi_{-(m-1/2)-1,n_{j}}$, we find that
\begin{equation}
\lim_{\mu\to-1/2}\frac{W(\{\gamma_\mu\})}{(4\mu+2)^N}
\propto W(\{\gamma\})\,.
\end{equation}
This relation is true for the case $i=1$, which implies that $\A_{2}\to\A_{2}^{\Omega}$ in the corresponding 
limit.  The general case is proved by induction.

\section{Wronskians (\ref{Interpoeg}), (\ref{Adlerkreineg}) }
\label{Apendix-Wronskian}
The  explicit form of the Wronskians  (\ref{Interpoeg})
and  (\ref{Adlerkreineg}),  which are used in the main text, is 
\begin{eqnarray}
\begin{array}{ll}
W(\psi_{\nu,2},\psi_{-\nu-1,2})=&(2\nu+1)e^{-x^2}\big(45 - 72 \nu +16 (-4x^6 + x^8) \\
& +8x^4(15 - 4 \nu (1 + \nu)) + \nu^2 (-7 +2 \nu (2 + \nu))\big),
\end{array}\\
\begin{array}{ll}
W(\psi_{\nu,2},\psi_{\nu,3})=&e^{-x^2} x^{
 3 + 2 \nu} \big(16 x^8 - 32 x^6 (5 + 2 \nu) + 
   24 x^4 (5 + 2 \nu)^2 - \\
  & 8 x^2 (3 + 2 \nu) (5 + 2 \nu) (7 + 
      2 \nu) +  (3 + 2 \nu) (5 + 2 \nu)^2 (7 + 
      2 \nu)\big).
\end{array}
\end{eqnarray}

\end{document}